\documentstyle[12pt]{article}

\begin{document}

\baselineskip=7mm

\begin{center}

{Many-body Systems Interacting via a 
Two-body Random Ensemble (I):
Angular Momentum distribution in the ground states}

{ Y. M. Zhao$^{a,c}$,
  A. Arima$^{b}$, and N. Yoshinaga$^a$}

\vspace{0.2in}
{ $^a$ Department of Physics, Saitama University, Saitama-shi, Saitama 338 Japan \\
$^b$ The House of Councilors, 2-1-1 Nagatacho, 
Chiyodaku, Tokyo 100-8962, Japan \\
$^c$ Department of Physics,  Southeast University, Nanjing 210018 China }

\date{\today}

\end{center}

\begin{small}

In this paper, we discuss the  angular momentum distribution
in the ground states of  
many-body systems  interacting via a  
two-body random ensemble.
Beginning with a few simple examples, a simple approach to predict
$P(I)$'s, angular momenta  $I$ ground
state (g.s.) probabilities,  of 
a few solvable cases, such as fermions in a small
single-$j$ shell and $d$ boson
systems, is given. This method is generalized
to predict $P(I)$'s of  more complicated cases,
such as even or odd number of fermions  
  in a large single-$j$ shell or  a many-$j$ shell,
 $d$-boson, $sd$-boson or $sdg$-boson systems,  etc.
By this method we are able to tell  
which interactions are essential to produce a sizable
$P(I)$ in a many-body system. 
The g.s.  probability
of maximum angular momentum $I_{max}$ is discussed.
An argument on the microscopic foundation of our approach,
and  certain matrix elements which are useful
to understand the observed regularities, 
are also given or addressed in detail.
The low seniority chain of 0 g.s. by using the same set of
two-body interactions is confirmed but it is  noted
that contribution to the total 0 g.s. probability beyond this chain
may be more important for even fermions in a single-$j$ shell. 
Preliminary results by taking a displaced  two-body random ensemble
are presented for the $I$ g.s. probabilities.

\end{small}

        
\newpage

\section{INTRODUCTION}

Many-body systems  interacting via  a scalar  two-body random matrix elements
are expected to have eigenstates which are (nearly)   
random superpositions of Slater determinants.
Among all the quantum numbers, the particle number  $n$  and the
total angular momentum $I$  of the states  are the only good quantum numbers.
Regularities exhibited by finite many-body systems
interacting  via random matrix elements
 provide an excellent tool to
study general features, which are 
independent of interactions,  of many-body systems \cite{Satula}. 
Therefore, robust regularities  
of many-body systems, if there are any,
are  both interesting and important.

Recently, Johnson, Bertsch, and Dean  discovered \cite{Johnson1} that
 the dominance of $0^+$ ground state (0 g.s.) of even 
fermion systems can be obtained by
using  a  two-body random ensemble (TBRE).
Further studies showed that
quasi-ordered spectra can be obtained  by using 
a two-body random ensemble \cite{Johnson2,Johnson3,Bijker}.   
 The 0 g.s. dominance  was soon confirmed in  $sd$-boson
systems \cite{Bijker,Casten}.
Therefore, this 0 g.s. dominance in even nucleon systems and
boson systems is  robust and insensitive
to the detailed  statistical properties of the random ensemble, suggesting
that the features of pairing arise 
from a very large ensemble of two-body matrix elements and might be 
independent of the specific 
character of the force.
An understanding of  this 0 g.s. dominance  is
important, because this observation
seems to be contrary to the traditional assumption  which
was taken previously. For example, in nuclear physics the 0 g.s.
dominance in even-even nuclei is explained as
a reflection of a strong pairing
associated with a strong short-range attraction between
identical nucleons.  Very recently, interesting
studies are performed to check whether the spectroscopy with
random and/or displaced random ensembles can simulate the
realistic systems \cite{Horoi,Zuker}.

There have been a few
efforts to understand this observation.  In Ref. \cite{Bijker1}, it was 
indicated that there is a correspondence between a large   
distribution width of $0^+$  states  and the  
0 g.s. dominance.  In Ref. \cite{Bijker}, it was suggested   
that for a system of interacting bosons the probability that 
the ground state has a certain value of the angular momentum 
is not really fixed by the full  
distribution  of eigenvalues, but rather by that of the lowest one. 
In \cite{zelevinsky}, Mulhall  et al. discussed the 0 g.s. dominance 
 fermions  in a single-$j$ shell  by using geometric chaoticity and
uniformly distributed random  interactions.
Kusnezov   discussed
the $sp$-boson systems by using random polynomials \cite{Kus}.
In Ref. \cite{Bijkerm,Bijkerx}, the authors
discussed the 0 g.s. dominance in  $sp$- and $sd$-boson systems
in terms of the mean field approach.
In Ref. \cite{Johnsonx} Kaplan et al.
studied the correlation between  
eigenvalues and spins of corresponding states in a few simple cases. 

For fermions in a  small single-$j$ shell,
it was  recently shown  \cite{Zhaox}
that the width of the energy distribution for each angular momentum $I$
is not the key  to understand the $0^+$ g.s. dominance, instead, the
coefficients $\alpha^J_{I   \beta \beta'} $
($n$-body matrix elements of two-body interaction
$A^{J \dagger} \cdot A^J$) with $ \beta = \beta'$
was suggested to provide a reasonable
explanation of the  distribution
of angular momentum $I$  g.s.,  where
$\alpha^J_{I   \beta \beta'}=$ $\langle n
 \beta I |$ $A^{J \dagger} \cdot A^J | n  \beta' I
\rangle$ with  $n$ being  the particle number,  
$\beta$   additional quantum number necessary to label the state,  and  
$I$ being the angular momentum of the state.
The $A^{J \dagger} \cdot A^J$ will be defined later. 
It was assumed \cite{Zhaox,Arima} that the off-diagonal matrix elements
$\alpha^J_{I   \beta \beta'} $ $(\beta \neq \beta')$ are
neglected as an approximation, which is based on an observation \cite{Zhaox}
 that they are very small compared with  diagonal  
 matrix elements  $\alpha^J_{I   \beta \beta} $  for $I > 0$
 states of even fermions in a small single-$j$ shell,  e.g., 
 $j=\frac{9}{2}, \frac{11}{2}$, if one uses a seniority conserved 
 basis \cite{Lawson}.
This  method   has  a disadvantage that it is not
applicable to more complicated
cases, e.g., fermions in a large single-$j$ shell, where the
off-diagonal matrix elements become also important.

These studies are interesting and important, and
have potentially impacted our understanding on the origin
of one of the most characteristic features of nuclear spectra.   
All these approaches, however,  address only    
simple or very specific  ($sp$ and $sd$ bosons, or fermions
in a  small  single-$j$ shell) cases.  
None of these arguments could explain the
regularities of angular momenta $I$ g.s. probabilities
such as those observed in \cite{Zhaox}. 
It is therefore  very desirable  to construct
 a {\it    universal}
approach to understand the 0 g.s. dominance of
even nucleon systems and $sd$-boson systems,  and  meanwhile,
to understand all the $I$ g.s. probabilities  (denoted as $P(I)$)
of very different systems including both even and
 odd number of  fermions in a single-$j$ or a many-$j$ shell. 
In this paper, we shall present an approach to
predict the $P(I)$'s of all types of systems
in a simple but universal procedure.

In this paper,
we use $G_J$'s be a set of Gaussian-type random numbers with  
a width being 1 and an average being 0:
\begin{equation}
  \rho(G_J) = \frac{1}{\sqrt{2\pi}} {\rm exp}(-G_J^2/2),~~~J=0,2,
  \cdots, 2j-1, \label{tbre}
\end{equation}
where the $G_J$'s define the two-body matrix elements of fermions in a 
single-$j$ shell  as follows:
\begin{eqnarray}
&& H = \sum_J G_J  A^{J \dagger} \cdot A^{J} \equiv  
\sum_J \sqrt{2J+1} G_J 
\left( A^{J \dagger} \times
\tilde{A}^J \right)^0, \nonumber \\
&&  A^{J \dagger} = \frac{1}{\sqrt{2}} \left( a_j^{\dagger} \times a_j^{\dagger}
 \right)^J, ~ ~
 \tilde{A}^J = - \frac{1}{\sqrt{2}} \left( \tilde{a}_j \times
 \tilde{a}_j \right)^J, ~ ~ G_J = \langle j^2 J|V| j^2 J \rangle.  \label{pair}
  \nonumber
\end{eqnarray}

For fermions in a many-$j$ shell  we use $G_J(j_1j_2j_3j_4)$'s in stead
of $G_J$'s.
For pure $d$ boson systems, there are only 3 independent  two-body
matrix elements, parameterized by  $c_0, c_2$ and $c_4$ \cite{Iachello}
which will be defined later.  For $sd$-boson systems
 we   shall also define the two-body hamiltonian separately.  
All the results, except a few cases which we shall note clearly, are
obtained by taking the two-body  matrix elements to be  
a  two-body random   ensemble (TBRE)
described by Eq.~(\ref{tbre}).

This paper is organized as follows:
In Sec. 2,
 We  begin with a few simple
but non-trivial cases$--$a $j=\frac{7}{2}$ shell with $n=3$ and  4, 
and $d$-boson systems with the number of bosons running from 3 to
48. In these cases,   eigenvalues are 
analytically given 
in terms of linear combinations of two-body
matrix elements. These   examples
 are very helpful to understand our simple approach
 proposed in this paper, and
also interesting because they are analytically given. 
In Sections 3 and 4,
we generalize our method  
and apply it to  both  even and odd numbers of  fermions in  
a large single-$j$ or a two-$j$ shell, $sd$- and
$sdg$-boson systems, where   this approach
continues to  work reasonably well.
For fermions in a single-$j$ shell it is found in this paper
that the $G_2=-1$ always produces $I=n$ ground state for even
$n$ fermions and $I=j-(n-1)/2$ ground state for odd $n$. 
In Sec. 5  we discuss the foundation of our simple approach, and
suggest  that the essential part of the  $I$ g.s. probability is related
to interaction $G_J$'s which produce angular momentum $I$ g.s.
if $G_J=-1$ and others are zero. This 
disproves a popular idea that the 0 g.s. dominance of even fermions
and bosons comes intrinsically
from the two-body nature of the interactions and might be
independent of the two-body interactions that used.  In Sec. 6, 
 we  address  
 the $I_{max}$ g.s. probability. 
For fermions in a single-$j$ shell. 
We present the eigenvalues  of the $I_{max}$ states
 with $n=$3, 4, 5, 6. This picture can be generalized
to explain a few features of the $I_{max}$ g.s. probabilities of
fermions in a two-$j$ shell.
In Sec. 7, 
 We shall check  the
previous statements of ``pairing" and   seniority ``chain" related
to the 0 g.s. dominance for fermions in a single-$j$ shell, where
seniority is well defined. We confirm the finding
of a low seniority chain suggested in \cite{Johnson2} but
note that the contribution to the 0 g.s. probability beyond
this chain may be more important.
A summary of this work will be given in Sec. 8. 
In appendix A we present our preliminary results  
  by using a displaced TBRE hamiltonian. In appendix B we give a simple
algorism to calculate diagonal matrix elements of fermions in a
single-$j$ shell.
In Appendix C, a few counter examples of the 0 g.s. dominance, and counter
examples of $I=j$ for odd fermins systems are also given.

\newpage

\section{A FEW SIMPLE SYSTEMS}

Generally speaking,   eigenvalues of a many-body system 
are   not linear in terms of two-body
  matrix elements. In some simple cases such as fermions in a  
single-$j$ shell  with $j \le$ 7/2   and  pure $d$-boson systems, 
however,  all eigenvalues are given in terms of
 linear combinations of two-body matrix elements.
For these  cases  an understanding of  the
$I$ g.s. probabilities using a concept of  shift  described
in Ref. \cite{Zhaox} is applicable: A state which has the largest
and/or the smallest coefficients for a given $G_J$ in the eigenvalues
is favored to be the ground state.
These examples will provide a useful 
clue to  obtain a universal approach.

\subsection{ $j=\frac{5}{2}$ and  $\frac{7}{2}$ shells }

Fermions in a single-$j$  ($j=\frac{5}{2}$ or  $\frac{7}{2}$) shell  are
  simple  but non-trivial cases. An analytical  relation between 
 eigen-energies $E_I$ and two-body  matrix elements $G_J$ 
are available.   

The eigen-energies of states with 3 fermions in a $j=\frac{5}{2}$ shell 
are given by 
\begin{eqnarray}
E_{3/2} &=& ~~~ ~~~~~~  {\bf \frac{15}{7} } G_2 +
             {\it   \frac{6}{7} } G_4, \nonumber \\
E_{5/2} &=&  {\bf  \frac{2}{3} } G_0  + ~  \frac{5}{6} G_2 ~  + 
                \frac{3}{2}  G_4, \nonumber \\
E_{9/2} &=& ~~~ ~~~~~~   {\it \frac{9}{14} } G_2  + 
              {\bf\frac{33}{14}} G_4.    \label{5/2}
\end{eqnarray}

The eigen-energies  $E_I$ of states with 3 fermions
in the  $j=\frac{7}{2}$ shell 
are as follows:
\begin{eqnarray}
E_{3/2} =& ~~~~~ ~~~~~~ ~~   \frac{9}{14} G_2 +
           {\bf \frac{33}{14}} G_4 +  {\it 0} G_6, \nonumber \\
E_{5/2} =& ~~~~~ ~~~~~~ ~~  {\bf \frac{11}{6}} G_2 +
             {\it \frac{2}{11} } G_4 +  \frac{65}{66} G_6, \nonumber \\
E_{7/2} =&  {\bf  \frac{3}{4} } G_0 +   \frac{5}{12} G_2 +
              { \frac{3}{4}} G_4 + \frac{13}{12} G_6,  \nonumber \\
E_{9/2} =& ~~~~~ ~~~~~~ ~~   \frac{13}{42} G_2 +
              \frac{150}{77} G_4 + \frac{49}{66} G_6,  \nonumber \\
E_{11/2} =&  ~~~~~ ~~~~~~  \frac{5}{6} G_2 +
              \frac{13}{22} G_4 + \frac{52}{33} G_6, \nonumber \\
E_{15/2} =& ~~~~~ ~~~~~~           {\it 0} G_2 +
              { \frac{15}{22}} G_4+  {\bf \frac{51}{22} } G_6.    \label{71/2}
\end{eqnarray}

  For four fermions in  a $j=\frac{7}{2}$ shell,
    each eigenstate    is labeled  using
seniority number  $v$ and angular momentum $I$.
The eigen-energies  $E_{I(v)}$ of states with 4 fermions  
are as follows:
\begin{eqnarray}
&E_{0(0)} = {\bf \frac{3}{2}} G_0 +  \frac{5}{6} G_2 +
              \frac{3}{2} G_4 + \frac{13}{6} G_6, \nonumber \\
&E_{2(2)} =\frac{1}{2} G_0 +  \frac{11}{6} G_2 +
              \frac{3}{2} G_4 + \frac{13}{6} G_6, \nonumber \\
&E_{2(4)} = ~~~~~ ~~~~~~            G_2 +
              {\bf\frac{42}{11}} G_4 + {\it \frac{13}{11}} G_6, \nonumber \\
&E_{4(2)} =\frac{1}{2} G_0 +  \frac{5}{6} G_2 +
    ~          \frac{5}{2} G_4 + \frac{13}{6} G_6, \nonumber \\
&E_{4(4)} =  ~~~~~~~~         {\bf \frac{7}{3}} G_2 +
                  ~~   {\it 1}   G_4 + \frac{8}{3} G_6, \nonumber \\
&E_{5(4)} =  ~~~~~~~~            \frac{8}{7} G_2 +
              \frac{192}{77} G_4 + \frac{26}{11} G_6, \nonumber \\
&E_{6(2)} =\frac{1}{2} G_0 +  \frac{5}{6} G_2 +
     ~         \frac{3}{2} G_4 + \frac{19}{6} G_6, \nonumber \\
&E_{8(4)} = ~~~~~    ~~~       {\it \frac{10}{21}} G_2 +
              \frac{129}{77} G_4 +  {\bf \frac{127}{33}} G_6.  \label{7/2}
\end{eqnarray}
In  Eqs.~(\ref{5/2}-\ref{7/2}),  
bold (italic) font is used
for coefficients  which are the largest (smallest) among all the
$I(v)$ states for a given $J$.   

We   rewrite Eqs.~(\ref{5/2}-\ref{7/2}) as  follows:
\begin{equation} 
E_{I(v)} = \sum_J \alpha^J_{I(v)} G_J  = \sum_J \alpha_{k}^J G_J,
  \label{alpha}
\end{equation}
where $k= I \beta \beta$ (=$I(v)$ in this
subsection), and  $\alpha_{k}^J$ satisfies  a sum-rule
 \cite{Lawson} 
\begin{equation}
\sum_J \alpha_{k}^J = \frac{1}{2} n(n-1).
\end{equation}
A method to calculate the above $\alpha^{J}_k$
for 4 fermions in a single-$j$ shell  is  given in terms
of 9-$j$ coefficients in appendix {\bf A}, where explicit expressions
are available for $v=0$ states. 
The $\alpha^{J_{max}}_{I_{max}}$   
 of fermions in a 
single-$j$ shell with $n=3$ to 6 is given in sec. 2.4.

By using a  TBRE hamiltonian described  by  
  Eq.~(\ref{tbre})
and the eigen-energies given by
Eqs.~(\ref{5/2}-\ref{7/2}), it is easy to obtain the 
 probability for each $I$ ground state ($I$ g.s.).
On the other hand, 
one can  predict   the  $I$ g.s. probability
 without running a TBRE hamiltonian. For
 example, the  exact  0 g.s. probability of 4 fermions in the
$j=\frac{7}{2}$ shell is determined by the
following integral:
\begin{eqnarray}
&& \int dG_0 \int dG_2  \int dG_4 \int dG_{6}
  \int dE_0
\int_{E_{0(0)}} dE_{2(2)}
\cdots  \int_{E_{0(0)}} dE_8
\nonumber   \\
&&  \delta \left(E_{0(0)} - \sum_J \alpha_{0(0)}^J G_J \right) \cdots
  \delta \left(E_{8(4)} -  \sum_J \alpha_{8(4)}^J G_J\right)
 \rho(G_0)  \rho(G_2)  \rho(G_4) \rho(G_6). \label{exact}
\end{eqnarray}

The $P(I)$'s of fermions in a single-$j$
shell  with $j=\frac{7}{2}, $ or $\frac{5}{2}$ 
  are given in Tables I-III. The row  ``TBRE"  corresponds to
results obtained by using  Eqs.~(\ref{5/2}-\ref{7/2}) and 1000
sets of  a TBRE hamiltonian. The
row ``pred1." corresponds to the probabilities  calculated
by integrals  such as  Eq.~(\ref{exact}) for the 0 g.s. probability
of 4 fermions in  a $j=\frac{7}{2}$ shell.
 Using Eqs.~(\ref{5/2}-\ref{7/2})
    the distribution  width,  $g_{I(v\beta)}$,
  of each state, is equal to
  $\sqrt{\sum_J \left( \alpha^J_{I(v)} \right)^2}$.
  These widths are listed in the last row of
 Tables I-III.

It is noticed that an understanding in terms of shifts  works well:
 a state with one or more largest (or the smallest) $\alpha_{I(v)}^J$ 
  has a very large probability to be the ground state or the
highest state. The $P(I)$'s of 
states without the largest and/or the
$\alpha_{I(v)}^J$ for a given $J$ are very small. A schematic argument
of such an observation was given in detail in \cite{Zhaox,Arima}. 

For 3 fermions in a  $j=\frac{5}{2}$ shell,  all   states
have the largest $\alpha_I^J$. Therefore, the above argument by using
  shifts  predicts that all  $P(I)$'s  
are  large, because  all the 3 states have the
largest $\alpha_I^J$ for a given $J$.    In  Table I,
  all states have indeed  large probabilities,
  which are obtained by running 1000 sets of a TBRE hamiltonian  and
shown in the row   ``TBRE", be the ground. 
The state  with $I=\frac{5}{2}$
  have only one  $\alpha_I^J$ which
 is the largest among different $I$ states,  while
 the other two states with $I=\frac{3}{2}$, $\frac{9}{2}$ 
 have both the largest and  the smallest  $\alpha_I^J$.
 Therefore,   $P(I=\frac{5}{2})$ is 
  smaller than $P(I=\frac{5}{2})$ and $P(I=\frac{9}{2}) $.

For 3 fermions in a $j=\frac{7}{2}$ shell,    states with   
$I=\frac{3}{2}$, $\frac{5}{2}$, $\frac{15}{2}$  
have both the largest and smallest $\alpha_I^J$, and the state
with $I=\frac{7}{2}$ has the largest $\alpha_I^J$ ($J=0$). 
Therefore, the above argument
predicts that the probabilities of
$I=\frac{3}{2}$, $\frac{5}{2}$, $\frac{7}{2}$ and  $\frac{15}{2}$   g.s. 
are  large.
  From Table II it is  noticed 
that all these  states have large probabilities to
be the ground states, but, 
the state $I=\frac{7}{2}$ state does not have  a coefficient  
 which is the smallest among $\alpha_I^J$'s with a given  $J$
 but different $I$, indicating
that the   $I=\frac{7}{2}$ g.s. probability
is   a bit smaller than those of $I=\frac{3}{2}$,
$\frac{5}{2}$, $\frac{15}{2}$.
 The   $I=\frac{9}{2}$
and $\frac{11}{2}$   states do not have coefficients
$\alpha_I^J$ which are  the largest or the smallest
among all the states. The predicted probabilities
of these two $I$ g.s. are small. Note that
it can be shown that the  $I=\frac{11}{2}$ cannot be
the ground state. All these features are confirmed by the  $P(I)$'s,
which are calculated  by running  1000 sets of a  TBRE hamiltonian,  
in the column ``TBRE".
                    
The situation is very similar in  the case
of 4 fermions in $j=\frac{7}{2}$ shell:
in each of the four states ($I(v)$= $0^+, 2^+(4), 4^+(4), 8^+$)
 which have large probabilities as
the ground state (and the highest state),
there are  coefficients $\alpha_{I(v)}^J$ which are  the
largest and/or the smallest for different $I(v)$.
 Other states which do not have the largest and/or the smallest
 $\alpha_I^J$ coefficients have very small probabilities
to be the ground state or the highest state. Refer to
the column   ``TBRE" in Table III. 

In Tables I-III, the $P(I)$'s obtained by running 1000 sets of 
a TBRE hamiltonian,  and those obtained  
by using integrals  such as  Eq.~(\ref{exact}) for the 0 g.s. probability
of 4 fermions for a $j=\frac{7}{2}$ shell, 
 are    well consistent
with each other.  Therefore,   
 the $I$ g.s. probabilities $P(I)$ 
in the above cases are explained  
in terms of   shifts, defined in Ref. \cite{Zhaox}.

However, the $P(I)$'s in
Eq. ~(\ref{exact}) is not yet within reach of a simple procedure
 by hand, and one has to
evaluate this integral  numerically. It is then very interesting and
important  to 
find  a simple method to evaluate the $I$ g.s. probabilities.

Let ${\cal N}'_I$   be   the number  
 of both the smallest and largest $\alpha_{I(v)}^J$
with a fixed $J$ for a certain $I$, the $I$ g.s. probability is approximately
given by ${\cal N}'_I /(N_m)$, where $N_m=2N-1$, $N$ is 
the number of two-body matrix elements.
The predicted $I$ g.s. probabilities by this   method
are also given in the row "pred2." of Tables I-III.
A reasonable agreement is obtained, 
  though there are small differences,
compared   with those obtained by running a  TBRE hamiltonian. 
Note that in the above examples, we use $N_m=2N-1$ because
all  $\alpha_{I(v)}^{J=0} (I \neq 0)$'s are 0 (there is
no smallest $\alpha_{I(v)}^{J=0}$).

\subsection{$d$-boson systems }

Similar to fermions in a small single-$j$ shell  
 ($j=\frac{5}{2}$ or  $j=\frac{7}{2}$), the  
 relation between the two-body matrix elements and the
 eigenvalues of  $d$-boson systems is also linear.

The two-body hamiltonian of a $d$-boson system is given by  
\begin{equation}
  H_d =  \sum_{l}  
  \frac{1}{2} \sqrt{2l+1} c_l \left( \left(d^{\dagger} \times
  d^{\dagger}  \right)^l \times \left( \tilde{d} \times \tilde{d}
  \right)^l \right)^0    \label{d-boson} 
\end{equation}

From Eq.~(2.79) of Ref.\cite{Iachello}, we have
\begin{equation}
E=E_0  + \alpha' \frac{1}{2} n_d (n_d-1) + \beta' \left[n_d (n_d+3) - v(v+3) \right]
+ \gamma' \left[ I (I+1) - 6n_d \right],       \label{eigen0}
\end{equation} 
where $E_0$ contributes only to binding energies
and not to excitation  energies, $n_d$ is the number of $d$ bosons.
Eq.~(\ref{eigen0}) can  be rewritten below:
\begin{equation}
E(v, n_{\Delta}, I)=E_0'(n_d) - \beta'  v(v+3) +  \gamma'   I (I+1).   \label{eigen}
\end{equation}

We cite Eq.~(2.82) of Ref. \cite{Iachello}:
\begin{eqnarray}
&&  \alpha' + 8 \gamma' = c_4, \nonumber  \\
&&  \alpha'-6  \gamma' = c_2, \nonumber  \\
&&  \alpha'+ 10 \beta' -12 \gamma' = c_0, \nonumber  
\end{eqnarray} 
which may be rewritten   below:
\begin{eqnarray}
&&  \alpha' = \frac{1}{7} (4c_2+3c_4), \nonumber  \\
&&  \beta'  =\frac{1}{70} (7c_0-10c_2+3c_4), \nonumber  \\
&&  \gamma' = \frac{1}{14} (-c_2 + c_4).   \label{coef}
\end{eqnarray}

Substituting these coefficients $\beta', \gamma'$ into
Eq.  (\ref{eigen}), and taking the two-body matrix elements
   $c_0, c_2$ and $c_4$ to be
the TBRE defined in Eq. (\ref{tbre}),
one obtains that {\bf only} $I$=0, 2, and $I_{max}$ ($=2n$) 
have sizable $I$ g.s. probabilities (other $I$ g.s.
probabilities are zero), which are shown in Fig. 1.
We notice following regularities of  
$P(I)$'s  vs. $n_d$. 

1. The $P(I_{max})$ is almost a constant (around 40$\%$)  for all
$n_d$ ($\le 4$); 

2. The $P(0)$ and $P(2)$ are periodical, with a period  $\delta(n_d)$=6.

3. All the $P(I_{max})$, $P(0)$  and $P(2)$  are near
to 0,  20$\%$, 40$\%$, or $60\%$.  The other $P(I)$'s  are
always zero. 

Below we explain these observations. 
From  Eq.~(\ref{eigen}) and Eq.~(\ref{coef}), we have
\begin{eqnarray}
  c_0=1, c_2=c_4=0: &&
E(v, n_{\Delta}, I)=E_0'(n_d) - \frac{1}{10}  v(v+3);
\nonumber  \\
  c_2=1, c_0=c_4=0: &&
E(v, n_{\Delta}, I)=E_0'(n_d) + \frac{1}{7}  v(v+3) - \frac{1}{14} I(I+1);
\nonumber  \\
  c_4=1, c_0=c_2=0: &&
E(v, n_{\Delta}, I)=E_0'(n_d) - \frac{3}{70}  v(v+3) + \frac{1}{14} I(I+1), 
\label{final}
\end{eqnarray}
where  $E_0'(n_d)$ is a constant for all states.
 Based on Eq. (\ref{final}), we obtain TABLE IV,
 which presents the angular momenta
giving the largest (smallest) eigenvalues
when $c_l=-1$ ($l=$0, 2, 4) and other parameters are   0 for
$d$ boson systems. In Table IV,
$\kappa$ is a  non-negative integer, and $n_d \ge 3$.
These angular momenta  appear periodically,  originating  from  the
reduction rule of U(5)$\rightarrow$SO(3). 
From TABLE IV, one notices again that
a certain $P(I)$ is large if 
one state with angular momentum $I$
 has the largest and/or the smallest $\alpha_{I\beta}^l$ (Eq.~(\ref{alpha}))
for a given $l$ ($l=0,2,4$).

Note that when  one searches for 
the smallest eigenvalue with $c_0=-1$ and
$c_2=c_4=0$  in case A of Eq.~(\ref{final}), one finds that 
many $I$ states are degenerate at the lowest value. 
Therefore, again, we  use  $N_m=3N-1$  in  predicting
the $P(I)$'s by the formula $P(I)={\cal N}_{I}/N_m$.
The results are well consistent,
without any exceptions,  with
those obtained by running a TBRE hamiltonian.
Take  $n=4$ case as an example,
${\cal N}_{I=0}=3$ and  ${\cal N}_{I=I_{max}}=2$.
We predict that 0 g.s. probability is 60$\%$ and
$I_{max}=8$ g.s. probability is 40$\%$ while all other
$I$ g.s. probabilities are zero. 
The 0 g.s. and $I_{max}$ g.s. probabilities given by
diagonalizing a TBRE hamiltonian are
60.7$\%$ and 39.3$\%$, respectively, and all other
$I$ g.s. probabilities are zero. 
Note that our predicted $P(I)s$ of $d$-boson systems are always 
consistent with those obtained by diagonalizing a TBRE hamiltonian
in the examples that checked: $n_d$=4 to 48. 
In another sentence, the distribution of the $P(I)$'s
can be explained satisfactorily by shifts produced by
the largest and/or smallest $\alpha_{k}^{l}$. 

\newpage 
\section{FERMIONS IN A LARGE  SINGLE-$j$ SHELL}

In this subsection we generalize the method proposed above,   
 and study the $P(I)$'s of fermions in  a 
single-$j$ shell  with particle number 
$n$=4, 5, 6, 7, 8. The explanation of our approach
will be addressed in Sec. 8. 

The procedures of our  approach to a general case are as follows. First,
set  one of the two-body matrix elements  
to be $-1$ and all  other interactions  to be zero.
Then one  finds which angular momentum $I$ gives
the lowest eigenvalue among {\bf all} the  
eigenvalues of the full shell model space.
Suppose that the number of independent two-body matrix elements 
is $N$, then the above procedure is iterated $N$ times. 
Each time  only one of  the $G_J$'s is set to be $-1$ while all the 
others are switched off. 
Next, among the  $N$ runs   one counts  how many times
(denoted as ${\cal N}_I$) of a certain angular momentum $I$ 
gives the lowest eigenvalue  among all the possible
eigenvalues. Finally, the probability of $I$ g.s. is 
  given by ${\cal N}_I/N_m \times 100 \%$.
Below we use $N_m=N$ (unless pointed out explicitly)    
 although the largest 
eigenvalues are equivalent to the lowest eigenvalues. The reason
is that these largest eigenvalues are usually (exactly or nearly) zero 
for many $I$ matrices, especially for a many-$j$ shell or a large ($j \ge
9/2$) single-$j$
shell. To have the ``rule" as simple  as possible, we shall 
use only the lowest eigenvalues  with
 one of  the $G_J$'s being set to be $-1$ and others  being switched off 
for fermion systems in  a large single-$j$ shell,
a many-$j$ shell, $sd$- and $sdg$-boson systems.
                                      
Fig. 2 presents a comparison between  the predicted $P(I)$'s 
 and those   obtained by diagonalizing a TBRE hamiltonian  
of fermions in a $j=\frac{9}{2}$ shell.
We present two cases:  $n=4$ (even) and $n=5$ (odd).
The agreements are good. Note that 
 such  agreements do  not deteriorate
(or become  even better) if one goes to cases of fermions in a 
larger single-$j$  shell  or a many-$j$ shell  where there are more
two-body matrix elements. 

Tables  V-IX give the angular momenta $I$ which produce the lowest 
eigenvalues for different two-body matrix elements and
particle numbers, with 
$G_J$ being $-1$ and others being  0,
and with  $n$ as much as possible.    In Table VI,
the number of ${\cal N}_0$ staggers with 
$j$ at a period of $\delta j$=3, which will certainly 
produce a staggering of the 0 g.s. probabilities with $j$ for 4 fermion
systems. Fig. 3 gives a comparison between the predicted
$P(0)$'s (open squares)  and those obtained by diagonalizing a
 TBRE hamiltonian  (solid squares) for $n=4$ and 6. 
It could be seen that a good    
agreement   is  obtained  for   fermions in  both a small
single-$j$ shell  and
a large single-$j$ shell. The predicted 0 g.s. probabilities 
exhibit  a  similar staggering  as  those obtained by 
diagonalizing a  TBRE hamiltonian.

 It is interesting to note  
 that   $P(0)$'s  can also be fitted by empirical formulae.
For example,  $P(0)$'s can be predicted by
\begin{eqnarray}
&& {\rm for} ~ n=4: 
P(0) = \frac{ \left[ (2j+1)/6 \right] + k }{j+\frac{1}{2}}
\times 100 \%,  k =
\left\{
\begin{array}{ll}
1       &  {\rm if}~ 2j = 3m   \\
0       &  {\rm if}~ 2j+1 = 3m   \\
-1      &  {\rm if}~ 2j-1 = 3m   
\end{array}  \right.;
\nonumber  \\   
&& {\rm for} ~ n=6: 
 P(0) = \frac{ \left[ (2j)/3 \right]}{j-\frac{1}{2}}  
\times 100 \%,
\end{eqnarray} 
where the $``\left[ ~~ \right]"$ means to take  the integer part. 
These empirical formulas are  
interesting because it presents a scenario without any calculations 
for very large-$j$ cases where  it would be too time-consuming to
diagonalize a  TBRE hamiltonian.  

A very interesting note of Tables V-IX is on the quadruple
matrix elements $G_2$ term.  It has been well known
for a few decades,  based on the seniority scheme, that the monopole pairing
interaction always gives
$I=0$ ground state for even fermion systems in a single-$j$ shell  
and $I=j$ ground state for odd number of fermions in a single-$j$ shell 
 when $G_0$ is set to be -1 and others 0.
However, little was known   about  
  the $G_2$  matrix elements in a single-$j$ shell.
The Tables V-IX show that
the quadruple pairing interaction corresponding to $G_2$ always gives
$I=n$ ground state for even fermion systems
and $I=j-(n-1)/2$ ground state for an odd number of fermions  
 when $G_2$ is set to be -1 and others 0.
A study of this observation based on pair approximation
is now in progress \cite{Zhao-preparation}.
 
\newpage

\section{ $sd$  and $sdg$ BOSONS, AND FERMIONS IN A MANY-$j$ SHELL}

Although  $sd$- and $sdg$-boson systems, and even and odd numbers of 
fermions in a many-$j$ shell  
are very different systems from the cases discussed above, 
our method is applicable. 
All features are explained   similarly.

The hamiltonian of a $sd$-boson system is as follows:
\begin{eqnarray}
H_{sd} = &&   H_d   +    e_{ssss}   \frac{1}{2}  (s^{\dagger}    s^{\dagger}) (ss)      
            + e_{sddd} \left(  \sqrt{\frac{1}{2} }
     (s^{\dagger}    d^{\dagger})  
  \left( \tilde{d}  \tilde{d}  \right)^2 +h.c. \right)
              \nonumber  \\
   &&        + e_{ssdd} \left(  \sqrt{\frac{1}{2} }
     (s^{\dagger}    s^{\dagger})  
  \left( \tilde{d}  \tilde{d}  \right)^0 +h.c. \right)
          + e_{sdsd}  
    \left(  (s^{\dagger}    d^{\dagger}) \times  
   ( s  \tilde{d} )  \right)^0,          \nonumber 
\end{eqnarray}
where $H_d$ is a two-body hamiltonian defined in
Eq. (\ref{d-boson}).

TABLE X presents the angular momenta which give the lowest energies
when one of the above parameters is set to be -1 and others 0.
We predict, according to Table X, that 
only $I=0$, 2, $2n$  g.s. probabilities are sizable, which is 
  consistent with  the previous observation \cite{Bijker,Bijkerx},
 that in   $sd$-boson systems  
 interacting  via  a two-body random ensemble  
 only $I=0$, 2, and $2n$ (maximum)
 have large probabilities to be the ground state, the 
  g.s. probabilities of other angular
 momenta are close to zero. 

Fig. 4 shows a comparison of the predicted
$P(I)$'s and those obtained by diagonalizing a TBRE hamiltonian
of $sd$-boson systems, with boson numbers ranging from
 6 to 16. It is seen that a good agreement is obtained.

The case of fermions in a many-$j$ shell is the most complicated.
Fig. 5 present a detailed comparison
of the predicted $P(I)$'s and those obtained by
diagonalizing a TBRE hamiltonian,  for fermions in
a two-$j$ ($j=\frac{7}{2}, \frac{5}{2}$) shell with
$n$=4 to 6.  The predicted $P(I)$'s 
 are reasonably consistent with those
obtained by diagonalizing a TBRE hamiltonian.

For fermions in a many-$j$ shell,   number of two-body matrix elements
is usually large.  In such cases, 
especially  in odd-fermion systems, there are ``quasi-degeneracy" problem
in counting ${\cal N_I}$:
 sometimes  the lowest eigen-value is  quite close to the second lowest
one when one uses $G_{J(j_1 j_2 j_3 j_4)}=-1$ and others 0. 
 For such two-body matrix elements,
one should actually introduce an additional ``rule"
in order to have a more reliable prediction.  
Namely, it is not appropriate to count ${\cal N_I}$ in a simple procedure.
In order to avoid confusions, however,
we did not modify the  way in counting ${\cal N_I}$ of such
cases  throughout this paper.  It is noted that
the $I=\frac{7}{2}$ in Fig. 5b) and $I=\frac{3}{2}$ in Fig. 5d)
belong to the case with ``quasi-degeneracy". Improvement 
of agreement between the predicted $P(I)$'s 
and those obtained by diagonalizing a TBRE hamiltonian can be achieved
by appropriately considering  the above ``quasi-degeneracy".

We have checked  two-$j$ shells such as  
$(2j_1, 2j_2)$=(5, 7), (5, 9), (11, 3), (11,5), (11,9) and 
(13,9) with $n=4, 5, 6$,
$sd$-boson systems with $n$ up to 17,  and $sdg$-boson systems
with $n=4, 5, $ and 6, 
and {\bf all} the  agreements are reasonably  good.

\newpage 
\section{A SCHEMATIC INTERPRETATION OF OUR APPROACH}

It is interesting and very important to know why our  simple approach
by numerical experiments can successfully produce 
$I$ g.s. probabilities which are in good
agreement with those obtained by diagonalizing a 
TBRE hamiltonian. 
In this subsection, we shall
provide a schematic explanation. A sound explanation may
 be much more sophisticated. 

As mentioned above, 
 the  relation between the eigenvalues and the
two-body matrix elements is usually not linear. However, 
eigenvalues are always linear in terms of two-body matrix elements 
in a ``local" space (explained below).
Namely,  within the  local space  
we may find linear relations between the eigenvalues and the two-body
matrix elements. Therefore, instead of studying the
effects of all the two-body matrix elements simultaneously,
we dismantle the problem into $N$ parts. In each part we focus
on only one term of two-body matrix elements.   As a schematic
interpretation of our method, we take $n$ fermions in
a single-$j$ shell as an example (It is easily
recoginzed that this explanation
is applicable to all complicated cases as well). Let us 
take a certain $G_J=-1$ and all $G_{J'}=0$ ($J' \neq J$), and 
diagonalize the two-body hamiltonian. Suppose that
 the   eigenvalues are  $E_{I \beta}^J$, and their 
corresponding  wavefunctions are 
\begin{equation}
\Phi (j^n, I \beta J) =  \sum_{K K'\gamma} \langle j^{n-2} (K \gamma) j^2 (K') | \}
j^n I \beta J \rangle \left( \Phi \left(j^{n-2} (K \gamma) \right) \times 
\Phi \left(j^2 (K') \right) \right)^I.
\label{linear}
\end{equation}
Now we introduce a small perturbation by
adding $\{\epsilon G_{J'}\}$.  $G_J=-1$ and
$\{\epsilon G_{J'}\}$ define our $(J/2)$-th local space
of two-body matrix elements.  Then
the  new eigen-energies are approximated by
\begin{equation}
\left( E_{I \beta}^J \right)' = E_{I\beta}^J
+ \epsilon \frac{n(n-1)}{2} \sum_{K K'\gamma J'}
\left[ \langle j^{n-2} (K \gamma) j^2 (K') | \}
j^n I \beta J \rangle \right]^2 G_{J'}.  
\end{equation}
Namely, the  $E_{I \beta}^J$'s are
linear in terms of $\{G_{J'} \}$ in the local space.
For  two-body matrix elements  
which are close to the above local space,
the angular momentum $I$,  which gives the lowest eigenvalue
among all $I'$'s with $G_J=-1$ and others zero, continues to
give the lowest eigenvalue.   This means that the a very large
part of full space of a
TBRE hamiltonian can be covered  by
the $N=j+\frac{1}{2}$ local subspaces
defined above, especially one uses
a TBRE hamiltonian which produce  a large
probability for small $|G_J|'$s.   This is the phenomenology
of our approach to predict the $I$ g.s. probabilities in this paper. 

A further rationale can be seen from the following analysis.  
To exemplify briefly, let us take 4 fermions in a 
single-$j$ shell with $j=\frac{17}{2}$. In Fig. 6a) we  set 
$G_{J_{max}} (J_{max} = 16)=-1$, and set all the other parameters are
taken to be the TBRE but with a factor $\epsilon$ multiplied.
We see that almost all cases of  the  g.s. belong  to
$I=I_{max}$ when  $\epsilon$ is small (say, 0.4).
If one uses $G_{J_{max}} (J_{max} = 16)= 1$, then the
$P(I_{max}) \sim 0$, which  means that the cases
of the TBRE with  $G_{J_{max}} < 0$
produce almost all the $I_{max}$ g.s. in a single-$j$ shell.   
In Fig. 6b) we present the  results of the same system  with 
$G_0$ being $-1$ and other $G_J$'s being the TBRE
multiplied by   $\epsilon$. It is seen similarly that  the 0 g.s.
is dominant for small $\epsilon$, and  that   
if we switch off all the interactions
which give  the $I=0$ lowest eigenvalue with  a certain $G_J$ being
 $ -1$ and others 0, then the
0 g.s. probabilities  will be very
small (such as $10\%$) or be close to zero ($\sim 2\%$). 
Therefore, by this method one readily find which interactions,
not only monopole pairing, are important
to favor the 0 g.s. dominance.
Previously, Johnson et al. noticed that the 0 g.s.
dominance is even independent
of monopole pairing \cite{Johnson1,Johnson2,Johnson3}.
It was not known, however,  whether  a certain two-body
matrix element is essential or partly responsible, and
how to find which interactions are essential, 
in producing the 0 g.s. dominance for a given system.

A shortcoming of the above explanation is as follows:
in our simple approach we set each $G_J=-1$ for each numerical 
experiment and find the angular momentum
of the lowest state. In most cases we obtain 
degenerate lowest states if we set $G_J=1$. Thus the
local space of $\{ G_J=1 + \epsilon G_{J'} (J' \neq J) \}$
is not considered according to the above explanation.
However, the good consistence of our predicted $I$ g.s.
probabilities with those obtained by diagonalizing
a TBRE hamiltonian seems to indicate that
the local space such as  $\{ G_J=1 + \epsilon G_{J'} (J' \neq J) \}$
is considered via certain procedures, namely,
the properties of local spaces
 defined by $\{G_J=-1 + \epsilon G_{J'}  (J' \neq J) \}$ 
are, more or less, enough to represent the main features of the full space, 
suggesting that the total $I$ g.s. probilities
of $G_J=-1$ local spaces might be symmetric
$as$ $a$ $whole$ to those of the $G_J=1$ local spaces.

\newpage
                                   
\section{The $I_{max}$ g.s. PROBALITIES}

For fermions in a  single-$j$ shell,  
the highest angular momentum (denoted as
$I_{max}$) state  was found to have a sizable probability to
be the  g.s. \cite{zelevinsky,Zhaox}. 
This  is explained  by the observation that 
 ${\cal N}_{I_{max}} =1$ always: 
One can  easily notice that  the eigenvalue of
$I=I_{max}$ state is the lowest when $G_{J_{max}} = -1$
and other parameters are switched off.  Because   
 ${\cal N}_{I_{max}} =1$, the predicted $I_{max}$ g.s. probabilities
 of fermions in  a single-$j$ shell  are 
$\frac{1}{N} = \frac{1}{j+1/2} \times 100 \%$, which are 
valid for all particle numbers (even or odd). It is
predicted  that the $I=I_{max}$ g.s. probabilities of
fermions in a single-$j$ shell  decrease  gradually
with $j$ and vanish  at a large $j$ limit; they will not saturate at a
sizable value with $j$.

Fig. 7a) shows the  $I_{max}$ probabilities for different particle numbers 
  in a  single-$j$ shell.  The agreement between
  $I_{max}$ g.s.  probabilities obtained by
 diagonalizing a TBRE hamiltonian  and those predicted by using  
  a simple $\frac{1}{N} \times 100\%$ is   good.

For $d$-boson systems, the $I_{max}=2n $ g.s. probabilities
for all $n$ are $\sim$40-42$\%$.
In Sec. 2.1.2, the predicted $P(I_{max})$'s are 
 ${\cal N}_{Imax}/5=40\%$, where ${\cal N}_{Imax} \equiv 2$.

For  $sd$-boson systems it was found in  
Ref. \cite{Bijker}   that the $I_{max}$ g.s. probabilities 
are  large, which  can be actually explained in the same way.  
 Among the  two-body matrix elements,
the interactions with  $c_4=-1$ and others being 0 produce 
the lowest eigenvalue for the $I_{max}=2n$ state.
The predicted $I=2n$  g.s. probability is 1/$N$=1/6=16.7$\%$,
which is independent of the boson number.  
This is consistent with  that obtained by
diagonalizing a TBRE hamiltonian ($\sim 15\%$).
Note that the  term $ (s^{\dagger} d^{\dagger}) (s d)$ gives degenerate
lowest eigenvalues   for many $I$ states when
$e_{sdsd}$ is set to be $-1$ and others are 0. Therefore,
we use 6 (instead of 7)  as the number of
independent two-body matrix elements, $N$.
The difference due to this minor modification is very  small, though.

For $sdg$-boson systems,
the predicted $I_{max}=4n$g.s probabilities is  
$1/N \sim 3.2\%$, where $N=32$. The $I_{max}$ g.s. probabilities that we 
obtain  by diagonalizing a TBRE hamiltonian are
$3.3\%$, $4.2\%$, $3.3\%$ for $n=$4, 5, 6, respectively.

The above 
argument of  $P(I_{max})$'s  can be generalized to more complicated
cases, such as fermions in a  many-$j$ shell,  bosons with
two or more different angular momenta (e.g., $sdg$ bosons).
 Let us firstly take a two-$j$ ($j_1, j_2$) shell. 
Similar to the above argument for fermions in a   
single-$j$ shell, it is predicted that the two angular momenta 
$I_{max}'=I_{max}(j_1^n)$ and $I_{max}(j_2^n)$
have   g.s. probabilities which 
are around or larger than 
$1/N \times 100\%$.
Here   $I_{max}(j^n)$ refers to the maximum among all  angular momenta 
of   states  constructed by   $j^n$ configurations. 
In another word, one can predict the
lower limit of these $I_{max}'$ g.s. probabilities.
Second, for a  boson system, e.g.,  a $sdg$-boson system, 
it is predicted that the $I=I'_{max}(d^n)=2n$  
  g.s. probability is
always larger than (or around)  
$1/32 \times 100\%=3.2\%$ (this probability obtained by
diagonalizing a TBRE hamiltonian is $\sim$12-13$\%$).

  Fig. 7b) presents the $I_{max}'=I_{max}(j_1^n),
  I_{max}(j_2^n)$  g.s. probabilities of fermions in a two-$j$ shell,
  obtained by diagonalizing a TBRE hamiltonian. They are compared with
the curve plotted using $1/N$. It is noticed that
the predicted lower limit of the $(I_{max})'$ g.s. probabilities
works quite well. 
It is worthy to mention that other $P(I)$'s  with $I$ very near 
$(I_{max})'$ are  almost zero (smaller than 1$\%$)
in these examples.

Now we study the eigenvalue of the $I_{max}$ state by
calculating the $\alpha_{I_{max}}^J$ of fermions in a single-$j$ shell.
 In doing so we present an argument
of an observation that  the $\alpha_{I_{max}}^{J_{max}}$
is always  lower than  other eigenvalues of all
 other states while $G_{J_{max}}=-1$ and others switched off.
 Equivalently, below we calculate
  $\alpha_{I_{max}}^{J_{max}}$ by setting 
  $G_{J_{max}}$=1 and others zero, which gives
the $\alpha_{I_{max}}^{J_{max}}$ the largest among  
 $\alpha_{I \beta \beta}^{J_{max}}$'s of all $I \beta$'s.

The calculation of $\alpha_{I_{max}}^{J_{max}}$ is
straightforward. 
By decoupling the two-body interaction operators and
 using analytical formulas of  Clebsch-Gordon coefficients, 
one can obtain all  $\alpha_{I_{max}}^{J}$'s.  
It is noticed easily  that
there are {\bf only} positive contributions in this state and 
 there are always cancellations in   all the
 other states when $G_{J_{max}}$=1 and others $G_J$'s switched off.
 The reason is as below:

The wavefunction of the $I_{max}$ state is known as
\begin{eqnarray}
| I_{max} M=I_{max} \rangle & = & 
      | j m_1, j m_2, \cdots j j_{m_n} \rangle \nonumber \\
& = & |j j, j (j-1), j (j-2), \cdots j (j+1-n) \rangle.  \nonumber 
\end{eqnarray}
All $I \neq I_{max}$ states can be constructed by a successive 
   orthogonalization with 
  those obtained by acting $J_{-}$ operator on  $| I_{max} M \rangle$ state.
It is easy to realize that 
 both negative sign and positive sign appear in
the wavefunctions of  $I \neq I_{max}$ states. 
  The coefficients in
$| I_{max} M  \rangle$  may be chosen to be   positive for all
$M$.  The $\alpha_I^{J_{max}}$ is given by a
summation of squares of  all possible   
couplings in the wavefunction. Therefore,
there is no cancellation in calculating  $\alpha_{I_{max}}^{J_{max}}$.
Cancellation appears if $I \neq I_{max}$. 

Below we list  some results for
$n=3$ to 6 fermions in a single-$j$ shell. 

1). For $n=3$:
\begin{eqnarray} 
\alpha_{I_{max}}^{J_{max}} &=& 2+ \frac{2j}{2(4j-3)}, \\ \nonumber 
\alpha_{I_{max}}^{J_{max}-2} &=& \frac{3 (2j-2)}{2(4j-3)}. 
\end{eqnarray}
                        
2). For $n=4$:
\begin{eqnarray}
\alpha_{I_{m ax}}^{J_{max}} & = & 3+\frac{2j (10j-11)}{2(4j-3)(4j-5)},
\nonumber \\
\alpha_{I_{max}}^{J_{max}-2} & = & 2-
\frac{ 2(2j)^3 - 11 (2j)^2 +  9(2j) + 15}{ (4j-3)(4j-5)(4j-7)}, \nonumber \\
\alpha_{I _{max}}^{J_{max}-4} &=& \frac{5(2j-3)(2j-4)}{2(4j-5)(4j-7)},
\nonumber \\
\alpha_{I_ {max}-2}^{J_{max}} &=& 3+\frac{2j (512j^3-2848j^2+5116j-2990)}{8
(8j-13)(4j-3)(4j-5)(4j-7)}.
\end{eqnarray}
 
3). For $n=5$:
\begin{eqnarray}
\alpha_{I_{max}}^{J_{max}}& = &4+ \frac{(2j)(8\cdot 2j -17)}{2(4j-3)(4j-5)}  \nonumber  \\
&& +\frac{5(2j)(2j-1)(2j-2)}{8(4j-3)(4j-5)(4j-7)}, \nonumber  \\
\alpha_{I_{ma x}}^{J_{max}-6} & = &  \frac{35 (2 j-4)(2j-5)(2j-6)}{8(4j-7)(4j-9)(4j-11)}.
\end{eqnarray}

4). For $n=6$:
\begin{eqnarray}
\alpha_{I_{max}}^{J_{max}}& = &3+ \frac{2j-3}{4j-3} 
 +\frac{148j^2-242j+60}{4(4j-3)(4j-5)} \nonumber \\
&& +\frac{2j(2j-1)(396j^2 -1482j + 1356)}{8(4j-3)(4j-5)(4j-7)(4j-9)}.
    \nonumber \\
\end{eqnarray}

\newpage

\section{CORRELATION BETWEEN THE GROUND STATES
OF FERMIONS IN A SINGLE-$j$ SHELL}

It is interesting to see whether there is any correlation
between different systems by the same two-body random interactions.
For this purpose the cases of fermions in a single-$j$ shell
is interesting because,
as shown in Sec. 2.2,  despite of the simplicity
 these cases  exhibit most
regularities of many-body systems interacting via a TBRE.

There are some evidences of correlation between ground states
of systems with  different particle number $n$ but
the same set of random interactions.
In this section we firstly report a correlation
of $I$ g.s. probabilities for fermions in a single-$j$ shell,
then examine another correlation discussed in \cite{Johnson2},
where this latter correlation was explained as reminiscence of
the general seniority scheme \cite{Talmi}.

The 0 g.s. probability of 4 fermions  in
a single-$j$ shell was found to fluctuate  periodically with $\delta_j=3$
\cite{zelevinsky,Zhaox}. We have checked $n=4$ up to $j=$33/2,
$n=6$ up to $j=27/2$ and $n=8$ up to $j=23/2$, which show that
the $0$ g.s. probabilities change synchronously  \cite{Johnson2}.
The $I=j$ g.s. probability of 5 fermions in a single-$j$
shell exhibits a similar pattern \cite{Arima}. 
In this paper  we  notice that   this  synchronous 
fluctuation appear at an interval
$\delta_j=3$ for 0 g.s. probability
of systems with $n=4$, 6, and 8 fermions
 and $I=j$ g.s. probability of odd $n$(=5,7) fermions   
 in a single-$j$ shell, but the origin of this correlation
 is not yet been available.

In a previous work \cite{Johnson2} another correlation
for fermions in a many-$j$ shell was
reported:  the pairing phenomenon seems to be 
favored simply as a consequence of the two-body
nature of the interaction. The ``pairing" here means that
there is a large matrix element between the
$S$ pair annihilation operator between the ground states
of a $n$ fermion system to a $n-2$, $n-4$ $\cdots$ system.
Below  we examine this ``pairing"
correlation for fermions in a single-$j$ shell, 
where the seniority quantum number $v$ is well defined.

First we see the simplest case: 4 and 6 fermions in
the $j=11/2$ shell. The 0 g.s. probability for $n=4$ and 6
is 41.2$\%$ and  66.4$\%$, respectively.
Among 1000 sets of the TBRE hamiltonian, 
364 sets  give 0 g.s. for both $n=4$ and 6 simultaneously.
This means that a TBRE hamiltonian in which $I=0$ is the ground
state for $n=4$ has a extremely large probability
(around 90$\%$) produces $I=0$ ground state for $n=6$.
The overlaps between the 0 g.s. 
of $n=6$ and the g.s. of $n=4$ coupled with an $S$ pair
for the same TBRE hamiltonian are in most cases around 0.8-0.9, while
those between the 0 g.s. of $n=4$ and that of $S$ pair
acting on the 0 g.s. of $n=6$ are in most cases around 0.9-1.0.
This strongly supports the finding in \cite{Johnson2}:
the $S$ pair annihilation operator takes
the 0 g.s. of $n$ fermions to that of $n-2$ fermions.
It is noted that the expectation values of seniority
for 0 g.s.  is, more or less, randomly distributed from 0 to 4,
in these cases.
Namely, there seems no bias for
very low seniority in the above calculation by
using a TBRE hamiltonian.

By examining   the 0 g.s.  
of 4 and 6 fermions
in the $j=11/2$ shell, one notices that the seniority  $v$'s 
of these g.s. are quite close. For $n=6$ 
there are strong seniority  mixings between states with
seniority $v=$0 and 4, but
no mixing between states with $v=6$ and 0 (or 4).  It is interesting to
note that no TBRE hamiltonian of $n=6$ for 0 g.s. with seniority 6
produces 0 g.s. of $n=4$. It is noticed that
the $I=0$ state with $v=6$ contributes $\sim 24\%$ to the total 
0 g.s. probability (66.4$\%$) for $n=6$. Roughly speaking,
in this small single-$j$ shell most of
the random two-body interactions which
produce 0 g.s.  with seniority $v\sim$0-4 of $n=6$ give
0 g.s. with a similar seniority for $n=4$, and in these cases the
picture given in \cite{Johnson2} is  appropriate. 

As for the $j=13/2$ shell  where there are strong 
mixings between states with $v=6$ and $v=$0 (or 4), 
0 g.s. probability is 22.3$\%$ for $n=4$   and  42.4$\%$ for
$n=6$. Among 1000 runs we obtain 13$\%$ sets of random interactions 
which give 0 g.s. for both $n=4$ and $n=6$. Among those cases  
which simultaneously produce 0 g.s. for $n=4$ and 6, 
the overlaps between the 0 g.s. 
of $n=6$ and the g.s. of $n=4$ coupled with an $S$ pair
for the same TBRE hamiltonian are in most cases around 0.6-0.9, while
those between the 0 g.s. of $n=4$ and that of $S$ pair
acting on the 0 g.s. of $n=6$ are in most cases around 0.8-1.0, 
indicating  a similar picture described in
\cite{Johnson2}.

Now we come to larger shells such as
$j=15/2$ to 23/2 with $n=$4, 6, and 8.
We concentrate on the  $j=15/2$ shell which is
  enough to demonstrate our points of view.
The 0 g.s. probability is 50.2$\%$, 68.2$\%$ and 32.1$\%$,
for $n=4, 6$ and  8, respectively. We have 31$\%$ (among
the 1000 runs) of the TBRE parameters which produce
0 g.s. for all $n=$4, 6 and 8, i.e., almost all those
TBRE parameters which produce 0 g.s. for $n=8$ produce
0 g.s. for $n=4$ and 6. It is interesting to note
that the expectation value of seniority $v$ of all  
31$\%$   0 g.s. of $n=4$, 6 and 8 systems 
has a very small probability larger than 4, indicating a similar
pattern given in \cite{Johnson2}.

On the other hand, it is noted that the  the fluctuation
of 0 g.s. probabilities for $n=4$, 6, and 8 fermions
in the same single-$j$ shell can be large. For example,
the 0 g.s. of 6 fermions in a $j$=15/2 shell is 68.2$\%$ while
that of 8 fermions is 32.1$\%$, which means that
more than 50$\%$ of the 0 g.s. for $n=6$ are not related to the
 chain in which the 0 g.s. of $n-2$ fermions can be obtained
by   annihilating one $S$ pair on that of $n$ fermions.
Fig. 8 presents a few cases of 4 or 6 fermions in a single-$j$ shell,
where no bias of a low seniority is observed, indicating that
the contribution to the total 0 g.s. probability beyond the
a seniority chain described in \cite{Johnson2} may be
more important in realistic systems. 
For example, there are only 14$\%$ of the TBRE hamiltonian parameters 
which produce 0 g.s. for both $n=4$ and 6 fermions in
a $j$=25/2, while the 0 g.s. probability of $n=6$ is 57.1$\%$. 

Therefore, we conclude that the finding  \cite{Johnson2} of
a chain of 0 g.s. which
is linked by $S$ pair for fermions in a  many-$j$ shell is also observed
frequently for even fermions in a single-$j$ shell, and that
this chain covers, however, only one part of the 0 g.s. and the contribution 
beyond this chain may be more important.

 It is noted that   no bias of low seniority
in the 0 g.s. is observed in our calculations by using
 a TBRE hamiltonian except that $n=8$ fermions in a $j=15/2$ shell,
 where no states with seniority 8 are observed in the $I=0$ ground states,
 i.e., most of 0 g.s. have expectation value of $v$ from 0 to 4.

\newpage

\section{ SUMMARY AND DISCUSSION}

To summarize,   we have presented in this paper
our understanding of regularities
of many-body systems interacting via  a  two-body random ensemble.

First,
beginning with simple systems 
in which the relation
between   eigenvalues and two-body matrix elements is linear, we propose an
integral to predict the  $I$ g.s. probabilities, $P(I)$'s. The properties
of $P(I)$'s are understood by the shifts,
or the largest and/or smallest eigenvalues with only one of the two-body
matrix elements switched on and others switched off.
This argument is further developed   
 to predict the $P(I)$'s by a simple formula $P(I)={\cal N}_I /N_m$, 
where ${\cal N}_I$ is the number of a certain angular momentum $I$
gives the lowest eigenvalue among all the possible
eigenvalues with   one of the two-body matrix elements being 
$-1$ and others being zero, and $N_m$ is taken as the number
of two-body matrix elements
except for   very few cases. 
 This method, as we show by    
a variety of very different systems, is  applicable to 
both (even or odd number of) fermion systems
(with both a single-$j$ shell and a many-$j$ shell)  and boson systems.
The agreement between the predicted $I$ g.s. probabilities
and those obtained by diagonalizing a TBRE hamiltonian  
  is good. It is noted that this method
predicts   the 0 g.s. probability, and on the same footing  it addresses
 other $I$ g.s. probabilities as well.
Therefore, we provide  in this paper a universal   
approach in studying the $I$ g.s. probabilities.

Next, we discuss the microscopic foundation of our simple approach by
defining a local space defined by
$G_J=-1 + \{\epsilon G_{J'} (J' \neq J) \}$. We show that
the angular momentum $I$ which gives the lowest eigenvalue
when $G_J=-1$ and others are zero continue to be the lowest
if $\epsilon$ is small.  One has the $I$ g.s. probability
around 70-90$\%$ even  $\epsilon$ is quite large (such as
0.5-1.0).  This could be a naive explanation of
the success of our approach, but
a sound explanation is not yet available. 

A  discovery of this work is that
 we are able to tell (by numerical experiments) 
 which interactions, 
not only the monopole pairing interaction,  are essential in favoring 
 0 g.s.  in both boson  fermions systems.  
For instance, 
interactions with $J=0, 6, 8, 12, $ and 22 give the 0 g.s.
dominance for 4 fermions in a $j=\frac{31}{2}$ shell.
This disproves a popular idea that the 0 g.s.
dominance essentially comes from the two-body nature
of the interactions and
might be independent of the form of the hamiltonian.

The simple $I_{max}$ g.s.
probabilities of fermions in a single-$j$ shell are
found,  for the first time,  to be determined by only the
number of the two-body matrix elements and
independent of particle number, and to follow a simply $1/N$ relation.
This phenomenon is explained by the fact that 
${\cal N}_{I_{max}}=1$  for fermions in a single-$j$ shell. 
A generalization of this regularity to fermions in a 
   many-$j$ shell   and
($sd$ and $sdg$) boson systems is
shown to work  well, too.    
Quite a few counter examples of the 0 g.s. dominance
in both fermions in a single-$j$ shell or a  multi-$j$ shell 
and bosons are found and explained.

One interesting note of the Tables V-IX is on quadruple
pairing interaction for fermions in a single-$j$ shell.
It is found   that $I=n$ ($j+1-n/2$)  give
the lowest eigenvalue when $G_2=-1$ and $G_{J\neq 2}=0$
for all even (odd) number of fermions in a single-$j$ shell that we 
have checked  in this paper.
 It is not known whether this observation 
is always correct, and what the origin might be if it is.

In this work we also studied the seniority
distribution for fermions in a single-$j$ shell where the
seniority number is well defined. In a pioneering work
\cite{Johnson2} it was claimed that the (low) seniority chain
is very important in the 0 g.s. dominance, namely, the
0 g.s. of $n$ fermions is related to 0 g.s. of
$n-2$, $n-4$ and etc  fermions via $S$ pairs approximately.
We confirm this phenomenon in this work. But we also note
that the 0 g.s. beyond this (approximate) seniority chain
may be more important.

It has  not yet understood at a more microscopic level
that   why the 
above ${\cal N}_0$ is  large for even fermion systems, or even  
more specifically,  why
there is a staggering on the ${\cal N}_0$ for even fermions  in  a
single-$j$ shell. 
Further consideration of this issue is warranted.

Finally, it is stressed 
that the $P(I)$'s discussed in this paper  and the
${\cal P}(I)$'s \cite{prc2}  (the 
probabilities of  energies averaged over all the 
states  for a fixed angular momentum $I$ 
being the ground states)  
are  different quantities. For even systems
the behavior of these two  are accidentally    
similar. For odd-$A$ systems, however, the ${\cal P}(I)$'s,
which were explained in terms of the randomness of two-body cfp's, 
 are very different from  ${P}(I)$'s. This 
 explicitly  demonstrates that
   the $I$ g.s. probabilities (and 0 g.s.
   dominance)  are not consequences of 
 geometric  chaoticity \cite{zelevinsky}. We show in this
 work that the 0 g.s. dominance is actually
 related to two-body matrix elements which have specific features.

{\bf Acknowledgement:} 
One of us (YMZ) is  grateful to Drs. S. Pittel, R. F. Casten, 
 Y. Gono,   Y. R. Shimizu, and  R. Bijker
for discussions and/or communications.
This work is  supported in part by the Japan Society
for the Promotion of Science
under contract No. P01021.

\newpage

{APPENDIX A ~~~ Many-body systems interacting via a displaced TBRE}

 While a TBRE is
symmetric to zero, an interesting question is that what  
  the results are if one  uses random interactions with {\bf only}
   positive, or negative sign, or random interactions which are not
symmetric to zero.  This  issue 
 is both interesting and
important because interactions in realistic systems, such as
nuclei, atoms etc, are not symmetric to zero. Below we
present preliminary results by using a displaced TBRE.

Let us  firstly consider two arbitrary ensembles
$\{ G_J' \}$ and $\{ G_J \}$,  which  
are related by a shift ${\cal C}$:
\begin{equation}
  G'_J = G_J + {\cal C},  ~~~J=0,2,
  \cdots 2j-1, \label{tbrex}
\end{equation}
where ${\cal C}$ is a constant. 

For fermions in a single-$j$ shell,   the results using the  
ensemble $\{ G_J' \}$ are exactly the same as those
obtained using  $\{ G_J' \}$ 
except a shift $\frac{n(n-1)}{2} {\cal C}$ on the
 eigen-energy of the ground state. Therefore, a 
 displacement of a TBRE is trivial in a single-$j$ shell. 

For fermions in  a  two-$j$ shell, however,  the role played by the displacement of
the TBRE is very complicated. Below we mention only two extreme situations:

1. 4 fermions in a  $(2j_1, 2j_2)=(11,3)$ shell, by using  
a TBRE, a TBRE+5, and a TBRE-5, 
the 0 g.s. probabilities are 60.8$\%$,
95.4$\%$, and 100.0$\%$, respectively. Here  
the 0 g.s.  is more pronounced if the TBRE is displaced to
either negative or positive.

2. 4 fermions in a   $(2j_1, 2j_2)=(13,9)$ shell, by using  
a TBRE, a TBRE+5, and a TBRE-5, 
the 0 g.s. probabilities are 44.8$\%$,
2.7$\%$, and 0.1$\%$, respectively. Here  
the 0 g.s.  is greatly quenched down (close to zero)
 if the TBRE is displaced to either negative or positive.
One observes a similar situation for 4 fermions
in a $(2j_1, 2j_2)=(7,5)$ shell.

Concerning the effect of
the shape of an ensemble, we  consider the TBRE and an ensemble
of uniformly changed random numbers between -1 and 1. 
The general features are quite similar.
Slight differences appear if one
multiply a  factor to each $G_J$, such as the RQE of [2]. 
There seems  no essential difference
between them, however.

\newpage

\begin{center}

APPENDIX~B~   Calculation of $\alpha^J_{I \beta \beta}$ for 4 fermions
in a single-$j$ shell 
\end{center}

 In this appendix, we calculate
  $\alpha^J_{I \beta \beta}$ for 4 fermions in a single-$j$ shell  by using
 9$j$ coefficients and coupled pair basis.  
                                                    
The normalized pair basis of a 4-fermion system 
is defined as follows:
\begin{equation}
|L_1 L_2: I \rangle_N = \frac{1}{ \sqrt{ N_{L_1 L_2 I} } }
 \left[ A^{L_1 \dagger} \times  A^{L_2 \dagger} \right]^I |0 \rangle
 = \frac{1}{ \sqrt{ N_{L_1 L_2 I} } } |L_1 L_2: I \rangle, \nonumber
\end{equation}
where
the subscript $N$ means that the state is normalized, and
$N_{L_1 L_2  I}$ is the overlap
$\langle L_1 L_2: I| L_1 L_2: I \rangle$, which is
given as follows:
\begin{equation}
N_{L_1 L_2 I} = 1 + \delta_{L_1 L_2} - 4  (2L_1+1) (2L_2+1)
 \left\{ \begin{array}{ccc}
j    & j  & L_1 \\
j    & j  & L_2 \\
L_1  & L_2 & I  \end{array} \right\}. \nonumber
\end{equation}

The matrix elements of two-body interaction within
normalized pair basis are given in Eq.~ (A.26) of Ref. \cite{Yoshinaga}:
\begin{equation}
 _N \langle L_1 L_2: I |
 \sqrt{2J+1}  \left[ A^{J \dagger} \times \tilde{A}^J \right]^0  |
L_1 L_2: I\rangle_N
= \frac{1}{ N_{L_1 L_2 I} } \sum_{R={\rm even}}  \left( U_{L_1 L_2 J R} \right)^2
 \label{matrix1}
\end{equation}
with
\begin{equation}
U_{L_1 L_2 J R} = \delta_{J L_1} \delta_{R L_2}
                + (-)^I \delta_{J L_2} \delta_{R L_1}
                - 4 \hat{L}_1  \hat{L}_2  \hat{J}  \hat{R}
 \left\{ \begin{array}{ccc}
j    & j  & L_1 \\
j    & j  & L_2 \\
J  & R & I  \end{array} \right\}. \nonumber
\end{equation}
Here $\hat{L}_1$ is a short hand notation of  $\sqrt{2L_1+1}$.

The seniority $v=$0 and 2 states are easy to
construct:
\begin{equation}
 \left\{ \begin{array}{cc}
|00:0 \rangle_0 =  |00:0 \rangle_N,   & v=0, \\
|I0:I \rangle_0 =  |I0:I \rangle_N,   & v=2   \end{array}~.  \right. \nonumber
\end{equation}
The subscript 0 means the state is a seniority-conserved state.
The seniority $v=4$ state is given by
\begin{equation}
|L_1 L_2: I \rangle_0 = \frac{1}{ \sqrt{1-\alpha^2} } |L_1 L_2: I \rangle_N
- \frac{ \alpha }{ \sqrt{1-\alpha^2} } |I 0: I \rangle_N,
\end{equation}
where $L_1 \neq 0, L_2 \neq 0$, and
$\alpha = _N\langle I 0: I| L_1 L_2: I\rangle_N$.
Suppose there are more than one seniority 4 states, say,
$|L'_1 L'_2: I \rangle_0 $ and $|L_1 L_2: I \rangle_0 $, one
should orthonormalize them to have an orthonormalized
basis.  We note here that in this case the
matrix elements $ _0\langle L_1 L_2: I|
\sqrt{2J+1}  \left[ A^{J \dagger} \times \tilde{A}^J \right]^0
|L_1 L_2: I \rangle_0$ and
 $_0 \langle L'_1 L'_2: I|
\sqrt{2J+1}  \left[ A^{J \dagger} \times \tilde{A}^J \right]^0
|L'_1 L'_2: I \rangle_0$ are not an invariant, their
summation is an invariant.

The $\alpha_0^J$ of seniority 0 state is as follows:
\begin{equation}
  _0 \langle 00: 0 | 
 \sqrt{2J+1}  \left[ A^{J \dagger} \times \tilde{A}^J \right]^0  |
00: 0\rangle_0  =
 \left\{ \begin{array}{cc}
2 \frac{2j-1}{2j+1}   & J=0, \\
8 \frac{2J+1}{4j^2-1},   & J\neq 0   \end{array}~.  \right. \nonumber  
\end{equation}

The 9$j$ coefficients are difficult to be further simplified
 unless one of them are 0.  
   The $\alpha_{I_{max}}^{J_{max}}$
is derived in another way in Sect.III.4.

\newpage

\begin{center}

APPENDIX~C~ A few interesting examples 
\end{center}

The 0 g.s. dominance 
was reported to be robust in even fermion systems and $sd$-boson systems.
Similarly, for fermions in a single-$j$ shell  it was believed that the 
$I=j$ g.s. probability is the largest \cite{zelevinsky}.
However, as shown in simple cases such as $d$-boson systems,
these dominances are {\bf not} always true.
Here we list a few counter examples that we noticed in our work.

1.   4 fermions in a $j=\frac{7}{2}$ shell, which was 
explained in terms of shifts   in \cite{Zhaox}; and
  4 fermions in a $j=\frac{13}{2}$ shell, which is understood in this
  paper by counting ${\cal N}_0$. In these two examples,
  $P(2)$'s are larger than $P(0)$.

2.   A two-$j$ shell  with $(2j_1, 2j_2)=(7,5)$ and $n=4$, where the 0 g.s.
probability is 21.5$\%$  and that of 2 g.s. is $34.7\%$; 

3.   two-$j$ shells with $(2j_1, 2j_2)$=(11,7), (13,9)  and $n=5$,
where  $j_1$ g.s. and $j_2$ g.s. probabilities
are   small ($\le 20\%$) while
$\frac{5}{2}$ g.s. probability is large ($> 30 \%$), which means that an
angular momentum which is not $j_1$ or $j_2$ may be favored
to be the ground state for odd number of fermion systems;

4.   $d$-bosons systems with $n\neq 6 \kappa$, while
     the 0 g.s. probability is less than 40$\%$
     (2 $\%$ or $\sim 36\%$)  while
     the $I_{max}$ g.s. probability is
     $\sim 42 \%$. Especially,
     the cases with $n=6 \kappa \pm 1$ ($\kappa$ is a natural number)
     are {\bf extreme} counter examples of the 0 g.s. dominance:  
      the 0 g.s. probabilities
     are very close to zero periodically.
     All regularities of  $P(I)$'s  of
     $d$-boson  systems are well understood in this paper.

\newpage
 
{TABLE I. The probability of each state to be  the ground
state and  distribution width of each eigen-energy
in the case of $j=5/2$ shell with
3 fermions. All the states are
labeled uniquely by their  angular momenta $I$.
The probabilities of the row ``TBRE" are obtained by 1000 runs 
 of a  TBRE hamiltonian, and those of  ``pred1."
 are obtained by integrals such as
 Eq.~(\ref{exact}) for $0^+$ state of $n=4, j=\frac{7}{2}$ case.
 The row "pred2." are obtained by the new  approach  
 proposed in this paper. 
 The  distribution  width,   $g_{I}$, of each eigen-energy,
 is listed in the last row.  }

\begin{tabular}{cccc} \hline \hline
$I$ & 3/2  & 5/2   & 9/2   \\  \hline
TBRE & $40.1\%$ & $23.7\%$ & $36.2\%$  \\
pred1. & 41.82$\%$ & 22.77$\%$ &  36.37$\%$  \\
pred2. & 40$\%$ & 20$\%$ &  40$\%$  \\
$g_{I}$ & 2.31 & 1.84 & 2.44  \\ \hline \hline
\end{tabular}

\vspace{0.3in}

{TABLE II. Same as Table I except that $j=7/2$ shell. }

\begin{tabular}{ccccccc} \hline \hline
$I$ & 3/2  & 5/2  & 7/2 & 9/2  & 11/2 & 15/2  \\  \hline
TBRE & $30.4\%$ & $29.7\%$ & $9.6\%$ & $4.4\%$ & $0.0\%$ &
$25.9\%$  \\
pred1. & 31.14$\%$ &27.26$\%$ &  9.70$\%$ & 3.69$\%$ &
0.00$\%$ &28.13$\%$  \\
pred2. & 28.6$\%$ &28.6$\%$ &  14.3$\%$ & 0 &
0 &28.6$\%$  \\  
$g_{I}$ & 2.44 & 2.09& 1.57 & 2.11 & 1.88 & 2.42 \\ \hline \hline
\end{tabular}

\vspace{0.7in}

{TABLE III. Same at Table II except that $n=4$. 
 All the eigenstates are uniquely
labeled by $I(v)$. }
                      
\begin{tabular}{ccccccccc} \hline \hline
$I(v)$ & 0(0) & 2(2) & 2(4) & 4(2) & 4(4) & 5(4) & 6(2) & 8(4) \\  \hline
TBRE & $19.9\%$ & $1.2\%$ & $31.7\%$ & $0.0\%$ & $25.0\%$ &
$0.0\%$ & 0.0$\%$  & 22.2$\%$ \\
pred1. & 18.19$\%$ & 0.89$\%$ & 33.25$\%$ & 0.00$\%$ &
22.96$\%$ & 0.00$\%$ & 0.02$\%$ & 24.15$\%$ \\
pred2. & 14.3$\%$ & 0$\%$ & 28.6$\%$ & 0$\%$ &
28.6$\%$ & 0 & 0$\%$ & 28.6$\%$ \\
$g_{I(v)}$ & 3.14 & 3.25& 4.12 & 3.45 & 3.68 & 3.62 & 3.64 &4.22  \\ \hline \hline
\end{tabular}

\newpage

{TABLE IV. The angular momenta which give the largest (smallest) eigenvalues
when $c_l=-1$  and other parameters are   0 for
$d$ boson systems.  } 

\begin{tabular}{cccccc} \hline  \hline
$n$   &  $c_0$(min) &  $c_2$(min) &  $c_2$(max) &  $c_4$(min) &  $c_4$(max)   \\  \hline
6$\kappa$  &      0      &  0  & $I_{max}$ & $I_{max}$ &0 \\
6$\kappa$+1&      2      &  2  & $I_{max}$ & $I_{max}$ &2 \\
6$\kappa$+2&      0      &  2  & $I_{max}$ & $I_{max}$ &2 \\
6$\kappa$+3&      2      &  0  & $I_{max}$ & $I_{max}$ &0 \\
6$\kappa$+4&      0      &  2  & $I_{max}$ & $I_{max}$ &2 \\
6$\kappa$+5&      2      &  2  & $I_{max}$ & $I_{max}$ &2 \\
    \hline  \hline
\end{tabular}

\vspace{0.4in}

{TABLE V. The angular momenta which give the lowest eigenvalues
when $G_J=-1$  and other parameters are   0 for
a  3-nucleon system in a single-$j$ shell. Here we use
a unit of $2I$, and
M refers to $I_{max}$. }

\begin{tiny}
\begin{tabular}{ccccccccccccccccc} \hline  \hline
$2j$ &  $G_0$ &  $G_2$ &  $G_4$ &  $G_6$ &  $G_8$ &  $G_{10}$ &  $G_{12}$
&  $G_{14}$ &  $G_{16}$ &  $G_{18}$ &  $G_{20}$ &  $G_{22}$ &  $G_{24}$
&  $G_{26}$ &  $G_{28}$ &  $G_{30}$   \\  \hline
5  &$2j$ &$2j-2$ &M &  &   &   & & & &  & & & & & & \\
7  &$2j$ &$2j-2$ &3 &M &   &   & & & &  & & & & & & \\
9  &$2j$ &$2j-2$ &$2j-6$ &9 &M  &   & & & &  & & & & & & \\
11 &$2j$ &$2j-2$ &$2j-6$ &3 &15 &M  & & & &  & & & & & & \\
13 &$2j$ &$2j-2$ &$2j-6$ &3 &9  &21 &M  & & &  & & & & & & \\
15 &$2j$ &$2j-2$ &$2j-2$ &5 &3  &15 &27 &M  & &  & & & & & & \\
17 &$2j$ &$2j-2$& $2j-2$& $2j-6$&3  &9  &17 &33 &M  &  & & & & & & \\
19 &$2j$ &$2j-2$& $2j-2$& $2j-6$&5  &3  &15 &23 &M-6 &M  & & & & & & \\
21 &$2j$ &$2j-2$& $2j-2$& $2j-6$&11 &3  &9  &17 &29 &M-6 &M  & & & & & \\ 
23 &$2j$ &$2j-2$&$2j-2$ &$2j-6$ &13  &5  &3 &11  &23  &35 &M-6 &M & & & & \\
25 &$2j$ &$2j-2$&$2j-2$ &$2j-6$ &15  &11  &3  &9 &17  &29  &41 &M-6 &M & & & \\
27 &$2j$ &$2j-2$&$2j-2$ &$2j-6$ &29      &13  &5   &3  &11&23 &35  &47  &M-6&M   & & \\
29 &$2j$ &$2j-2$&$2j-2$ &$2j-2$ &$2j-6$  &15  &11  &3  &9 &17 &29  &41  &53 &M-6 &M & \\
31 &$2j$ &$2j-2$&$2j-2$ &$2j-2$ &$2j-6$  &21  &13  &5  &3 &11 &23  &35  &47 &59 &M-6 & M 
\\   \hline  \hline
\end{tabular}
\end{tiny}

\newpage

{TABLE VI. The angular momenta which give the lowest eigenvalues
when $G_J=-1$  and other parameters are   0 for
   4 fermions in a single-$j$ shell.  }

\begin{footnotesize}
\begin{tabular}{ccccccccccccccccc} \hline  \hline
$2j$ &  $G_0$ &  $G_2$ &  $G_4$ &  $G_6$ &  $G_8$ &  $G_{10}$ &  $G_{12}$
&  $G_{14}$ &  $G_{16}$ &  $G_{18}$ &  $G_{20}$ &  $G_{22}$ &  $G_{24}$
&  $G_{26}$ &  $G_{28}$ &  $G_{30}$   \\  \hline
7  & 0 &4 &2 &8 &   &   & & & &  & & & & & & \\
9  & 0 &4 &0 &0 &12 &   & & & &  & & & & & & \\
11 & 0 &4 &0 &4 &8  &16 & & & &  & & & & & & \\
13 & 0 &4 &0 &2 &2  &12 &20 & & &  & & & & & & \\
15 & 0 &4 &0 &2 &0  &0  &16 &24 & &  & & & & & & \\
17 & 0 &4 &6 &0 &4  &2  &0  &20 &28 &  & & & & & & \\
19 & 0 &4 &8 &0 &2  &8  &2  &16 &24 &32 & & & & & & \\
21 & 0 &4 &8 &0 &2  &0  &0  &0  &20 &28 &36 & & & & & \\
23 & 0 &4 &8 &0 &2  &0  &10 &2  &0  &24 &32 &40 & & & & \\
25 & 0 &4 &8 &0 &2  &4  &8  &10 &6  &0  &28 &36 &44 & & & \\
27 & 0 &4 &8 &0 &2  &4  &2  &0  &0  &4  &20 &32 &40 &48 & & \\
29 & 0 &4 &8 &0 &0  &2  &6  &8  &12 &8  &0  &24 &36 &44 &52 & \\
31 & 0 &4 &8 &0 &0  &2  &0  &8  &14 &16 &6  &0  &32 &40 &48 & 56 
\\   \hline  \hline
\end{tabular}
\end{footnotesize}

\vspace{0.6in}

{TABLE VII. Same as TABLE V except that $n=5$.  } 

\begin{tabular}{cccccccccccc} \hline  \hline
$2j$ &  $G_0$ &  $G_2$ &  $G_4$ &  $G_6$ &  $G_8$ &  $G_{10}$ &  $G_{12}$
&  $G_{14}$ &  $G_{16}$ &  $G_{18}$ &  $G_{20}$   \\  \hline
9  &$2j$ &$2j-4$ &9 &3 &$I_{max}$   &   & & & &  & \\
11 &$2j$ &$2j-4$ &5 &$2j$& 5  & $I_{max}$   & & & &  & \\
13 &$2j$ &$2j-4$ &5 &$2j$& 7   & 5 & $I_{max}$ & & &  & \\
15 &$2j$ &$2j-4$&7 &9 & $2j$  & 7 &31& $I_{max}$ & &  & \\
17 &$2j$ &$2j-4$&9 &5 & $2j$  & 11&9 &41 & $I_{max}$ &  & \\
19 &$2j$ &$2j-4$&11&5 & 13  & $2j$&13&5 &51 & $I_{max}$ & \\
21 &$2j$ &$2j-4$&19&5 & 7   & $2j$&15&$2j$ &$2j$ & 61 & $I_{max}$
\\   \hline  \hline
\end{tabular}

\newpage

\vspace{0.6in}

{TABLE VIII. Same as TABLE V  except that $n=6$.  } 

\begin{tabular}{cccccccccccc} \hline  \hline
$2j$ &  $G_0$ &  $G_2$ &  $G_4$ &  $G_6$ &  $G_8$ &  $G_{10}$ &  $G_{12}$
&  $G_{14}$ &  $G_{16}$ &  $G_{18}$ & $G_{20}$ \\  \hline
11 &0 &6 &4 &0 & 0  & $I_{max}$ & & & &  & \\
13 &0 &6 &4 &0 & 4  & 4 & $I_{max}$& & &  & \\
15 &0 &6 &0 &6 & 0  & 0 & 0 & $I_{max}$ & & &  \\
17 &0 &6 &0 &0 & 0  & 4 & 0 & 16 & $I_{max}$ &&   \\
19 &0 &6 &0 &6 & 0  & 0 & 0 & 4  & 22 & $I_{max}$&  \\
21 &0 &6 &10&6 & 0  & 0 & 0 & 0  & 0 & 28 & $I_{max}$  
\\   \hline  \hline
\end{tabular}

\vspace{0.6in}

{TABLE IX. Same as TABLE V except that $n=7$.  } 

\begin{tabular}{ccccccccccc} \hline  \hline
$2j$ &  $G_0$ &  $G_2$ &  $G_4$ &  $G_6$ &  $G_8$ &  $G_{10}$ &  $G_{12}$
&  $G_{14}$ &  $G_{16}$ &  $G_{18}$  \\  \hline
13 &$2j$ &$2j-6$ &11 &1 & 11  & 5 & $I_{max}$& & &   \\
15 &$2j$ &$2j-6$ &13 &3 & 3   &$2j$ & 9 & $I_{max}$ & &   \\
17 &$2j$ &$2j-6$ &7 &$2j$ & $2j$  & 13 & 9 & 23 & $I_{max}$ &   
\\   \hline  \hline
\end{tabular}

\vspace{0.6in}

{TABLE X. Same as TABLE V except that $sd$-boson systems.
The matrix element  corresponding to $e_{sdsd}$ are omitted because
it always presents degenerate levels for many $I$ states. }

\begin{tabular}{ccccccc} \hline  \hline
$n$ &  $e_{ssss} $ &  $e_{sddd}$ &  $e_{ssdd}$ &   $c_0$ &  $c_2$ &  $c_4$
  \\  \hline
6 & 0 &0 &0 &0 & 0  & $I_{max}$    \\
7 & 0 &0 &0 &2 & 2  & $I_{max}$    \\
8 & 0 &0 &0 &0 & 2  & $I_{max}$    \\
9 & 0 &0 &0 &2 & 0  & $I_{max}$    \\
10& 0 &0 &0 &0 & 2  & $I_{max}$    \\
11& 0 &0 &0 &2 & 2  & $I_{max}$    \\
12& 0 &0 &0 &0 & 0  & $I_{max}$    \\
13& 0 &0 &0 &2 & 2  & $I_{max}$    \\
14& 0 &0 &0 &0 & 2  & $I_{max}$    \\
15& 0 &0 &0 &2 & 0  & $I_{max}$    \\
16& 0 &0 &0 &0 & 2  & $I_{max}$    \\
   \hline  \hline
\end{tabular}

\newpage

{\bf Figure captions}: 

FIG. 1 ~~The $I$ g.s. probabilities of $d$ bosons. The boson number $n$ 
runs from  4 to 44. Only states with  $I=$0, 2, and $I_{max}=2n$
are possible to be the ground. The 0  g.s., 2  g.s. and $I_{max}=2n$  g.s.
probabilities are very near to 0, 20$\%$, 40$\%$ or 60$\%$.
The $P(0) \sim 0$    in $d$-boson systems
 with $ {n_d} = 6 \kappa \pm 1$. 
 The predicted $P(I)$'s (open squares) are well
 consistent with those (solid squares)
 obtained by diagonalizing a TBRE hamiltonian.
All  regularities are explained by the reduction rule of
$U(5) \rightarrow O(3)$.

\vspace{0.2in}

FIG. 2 ~~  The $P(I)$'s  of fermions in a single-$j$ shell.
The solid squares are  $P(I)$'s obtained by diagonalizing a TBRE hamiltonian and the 
open squares are   $P(I)$'s  predicted by the approach proposed
in this paper. ~
a) $j=\frac{9}{2}$ with 5 fermions; 
b) $j=\frac{9}{2}$ with 4 fermions.
Good agreements are obtained for both even and odd-A
cases. 

\vspace{0.2in}

FIG. 3 ~~ The $P(0)$'s of fermions in a    
 single-$j$ shell. Solid squares 
 are obtained from 1000 runs of a TBRE hamiltonian.  
The   open squares  are predicted $P(0)$'s in this paper. ~  
a)  $n=4$; ~ b)   $n=6$. 
solid triangles in a) are obtained from  an empirical 
formula, Eq. (15).  

\vspace{0.2in}

FIG. 4 ~~  The $P(0)$, $P(2)$ and $P(I_{max}$) of $sd$-boson systems.
 Solid symbols  are $P(I)$'s obtained from 1000 runs of a TBRE hamiltonian.  
Open symbols  are $P(I)$'s predicted in this paper.  
Only $I=0$, 2, $I_{max}$  g.s. probabilities are
included. All other $P(I)$'s obtained by
diagonalizing a TBRE hamiltonian are close to zero, and
the predicted $P(I)$'s  are zero.

\vspace{0.2in}

FIG. 5 ~~ Fermions in a two-$j$ shell with  ($j_1, j_2$)
= $(\frac{7}{2}, \frac{5}{2})$. $n=$4, 5, 6, 7
in a), b), c) and d), respectively. Solid squares are
obtained by 1000 runs of a TBRE  hamiltonian and open squares are
  predicted by the method proposed in this paper.

\vspace{0.2in}

FIG. 6 ~~
 4 fermions
in a $j=\frac{17}{2}$ shell. ~   
a)  ~The $I_{max}$ g.s. probabilities with $G_{16}$=$\pm 1$
($J_{max}$=16) and
all other $G_J$ being  a  TBRE multiplied by $\epsilon$; 
b)  ~The $0$ g.s. probabilities with $G_{0}$=$\pm 1$
 and all other $G_J$ being a TBRE multiplied by $\epsilon$.
Refer to text for details. 

\vspace{0.2in}

FIG. 7 ~~ The $I_{max}$ g.s. probabilities.
~ a) The $I_{max}$ g.s.
probabilities of fermions in a single-$j$ shell, which are obtained
by diagonalizing a TBRE hamiltonian   (1000 runs). 
Solid squares are plotted by  $\frac{1}{N}\times 100 \%$,
$N=j+\frac{1}{2}$.
It is noted that the $I_{max}$  g.s. probabilities follow
a  $\frac{1}{N}\times 100 \%$ relation  well, and are  
 independent
of particle number, $n$.
 ~ b)  The $(I_{max})'=I_{max}(j_1^n)$ and  $I_{max}(j_2^n)$
  g.s. probabilities of fermions in  two-$j$
shells,
 obtained by 1000 runs of
a TBRE hamiltonian.
  The lower limit of the
 $(I_{max})'$  g.s. probabilities are predicted  
  $\frac{1}{N} \times 100\%$    
(solid squares).  The $I'_{max}$ g.s.
probabilities are reasonably consistent with the predictions.

\vspace{0.25in}

FIG.~8 ~~ The seniority distribution in the angular momentum $I=0$
ground states. No bias of low seniority is observed in these
4 and 6 fermions in a single-$j$ shell, which indicates that
the contribution to the total 0 g.s. beyond a low seniority chain
may be more important. 

\vspace{0.25in}

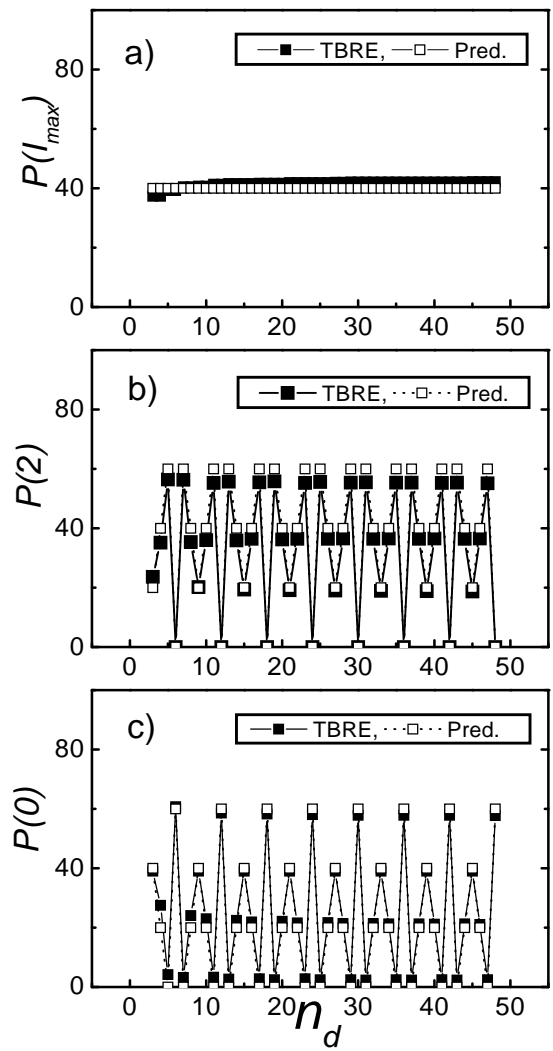
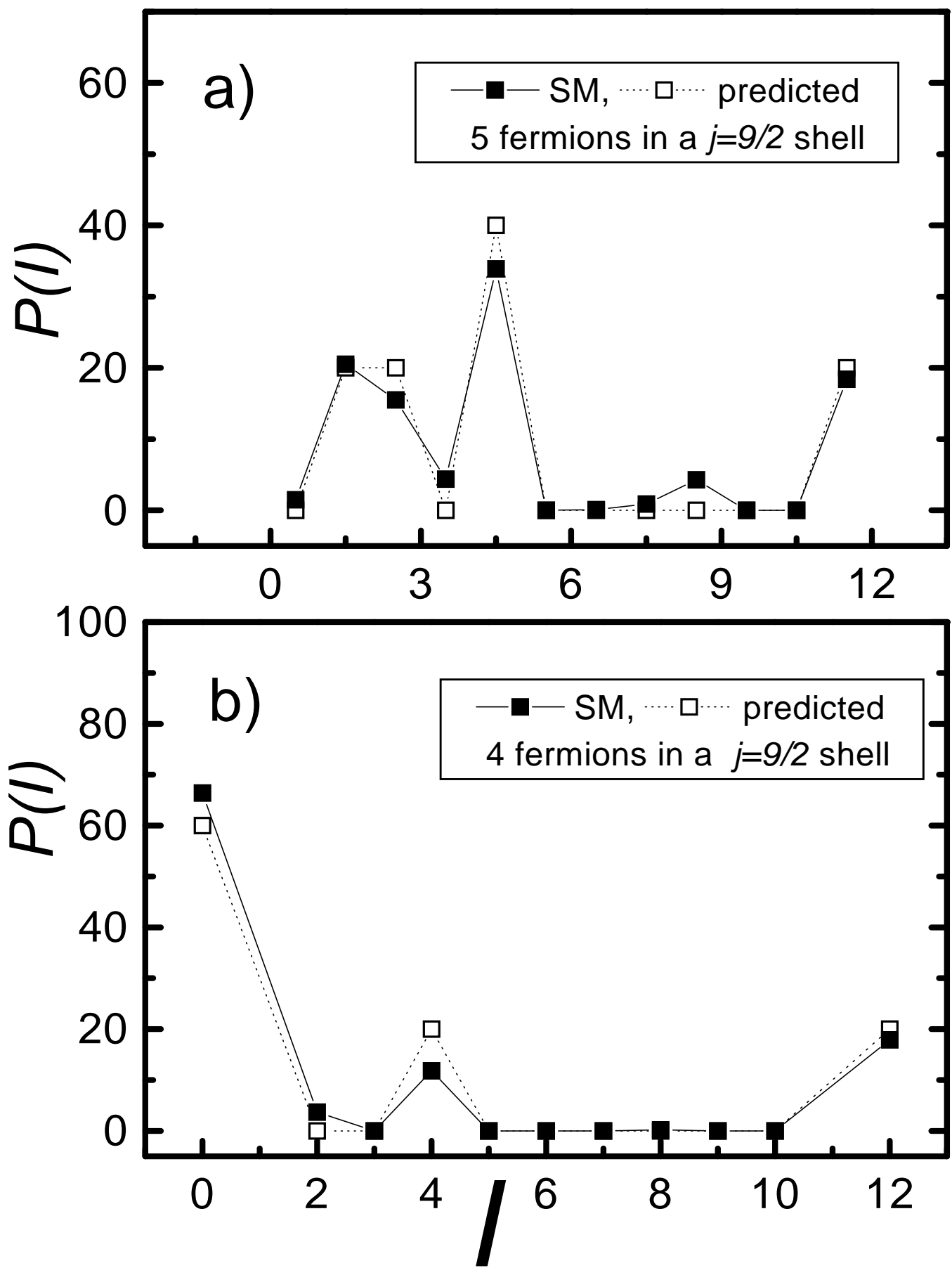
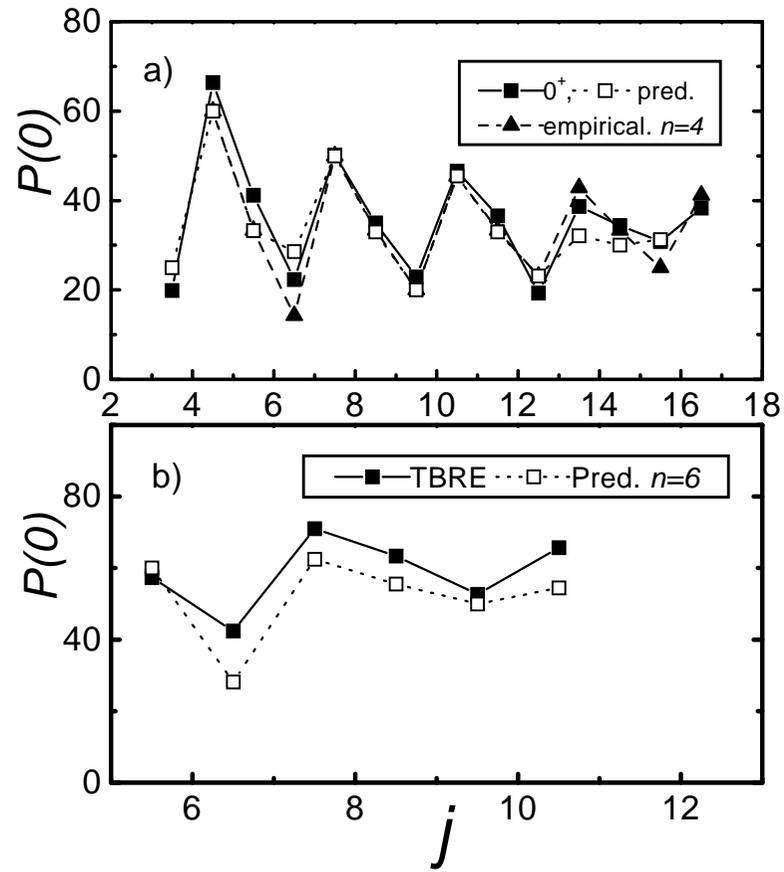
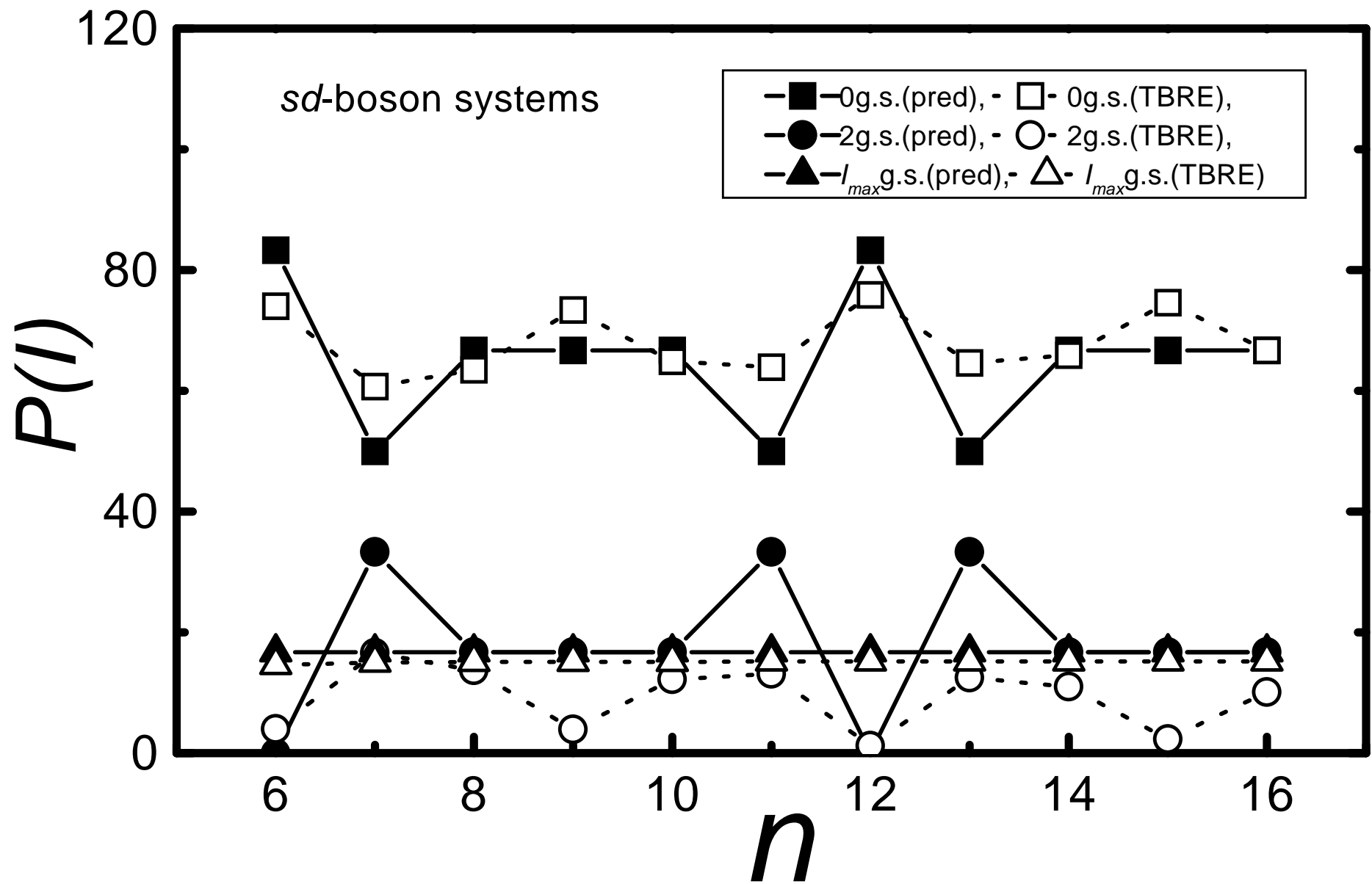
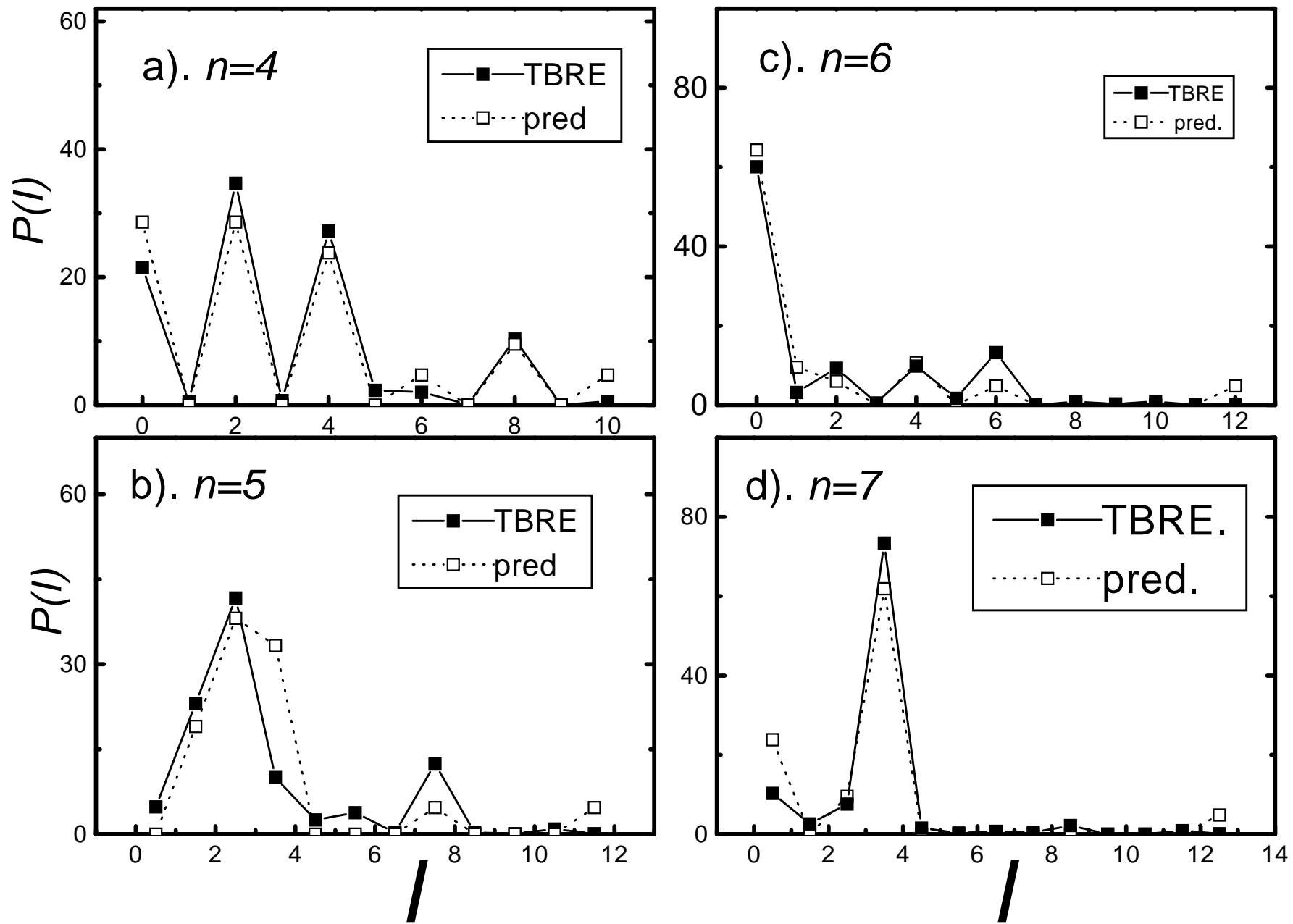
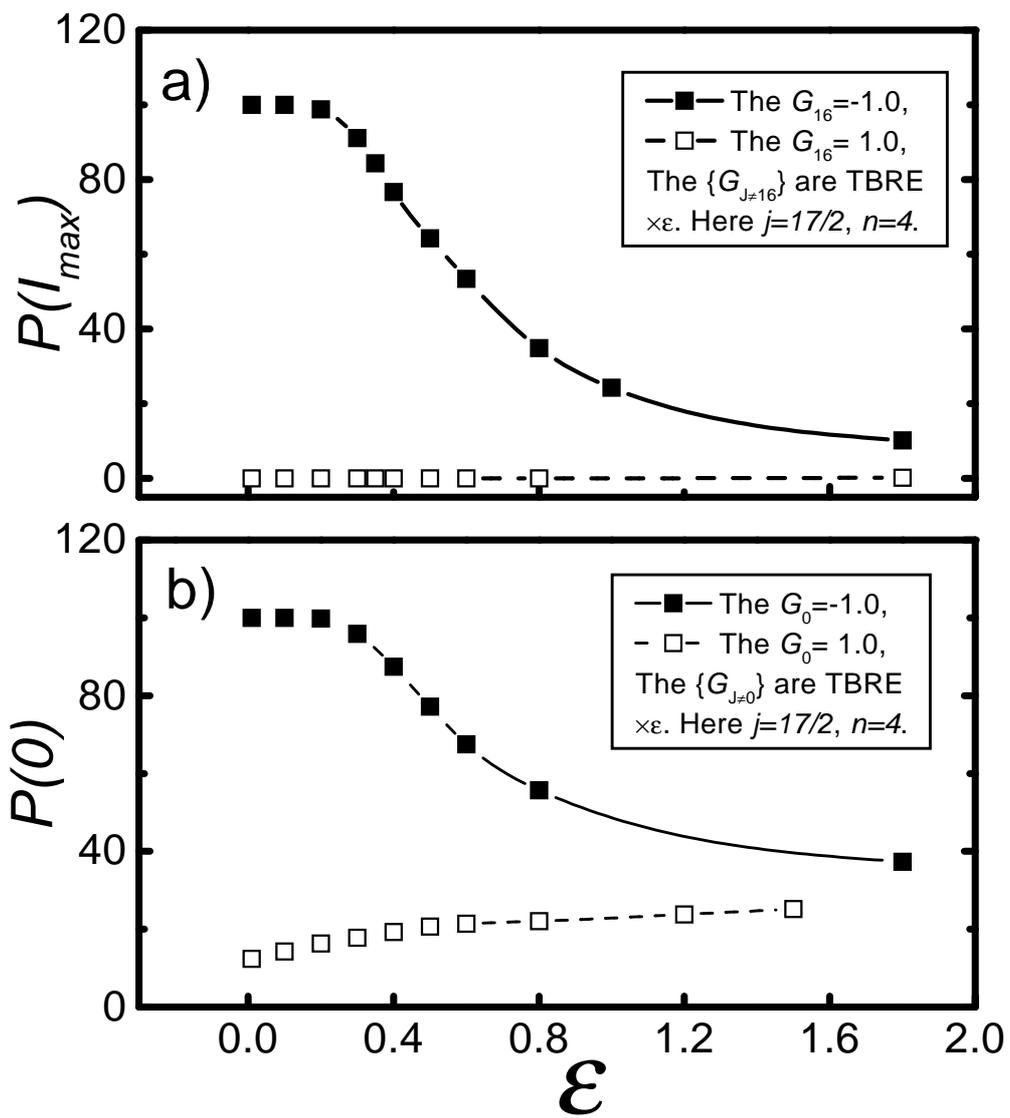
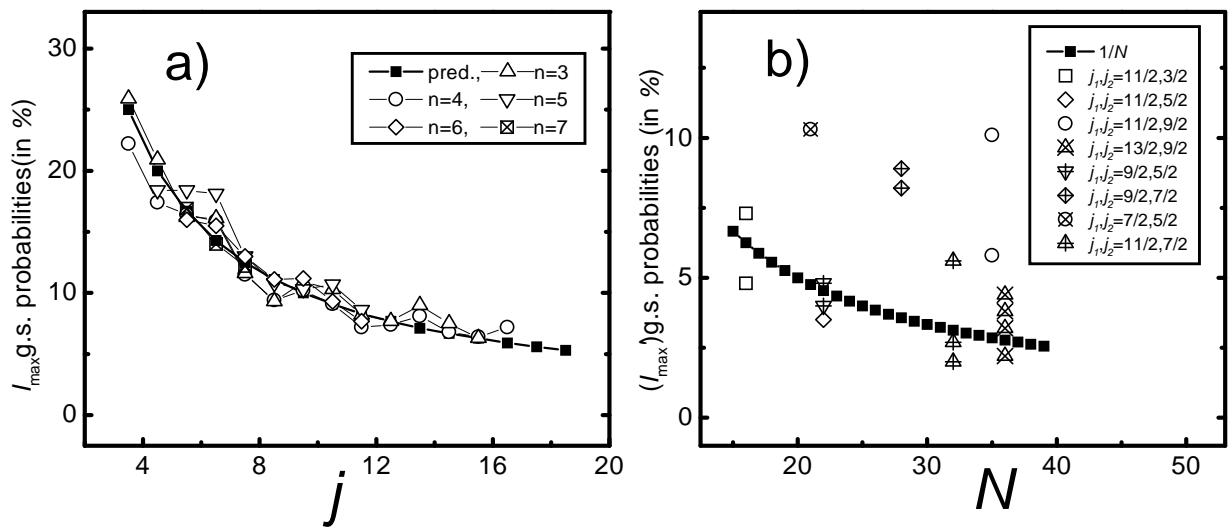
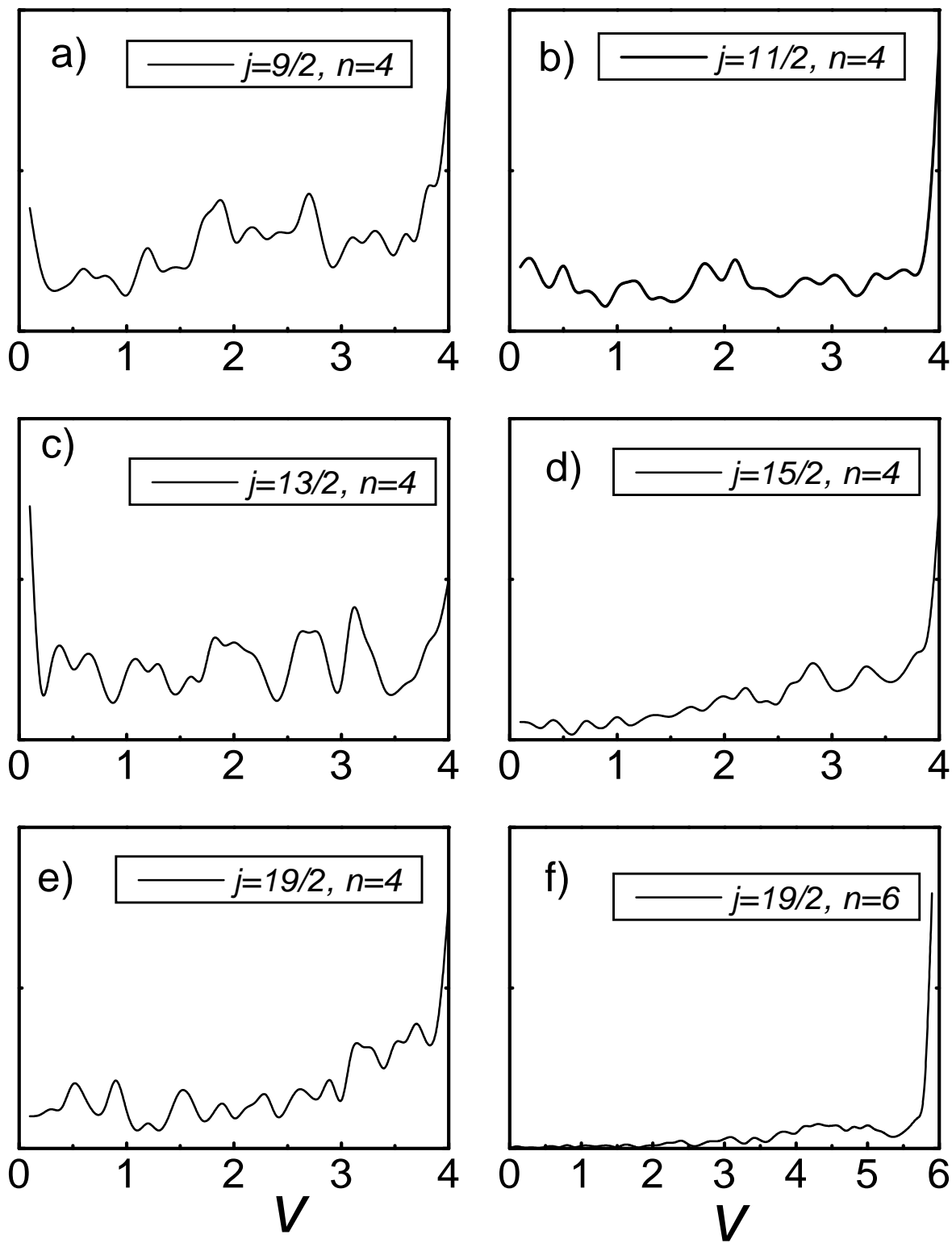
\end{document}